\documentclass[fleqn,usenatbib]{mnras}

\usepackage{newtxtext,newtxmath}
\usepackage[T1]{fontenc}
\usepackage{ae,aecompl}

\usepackage{mathrsfs}

\usepackage{graphicx}    % Including figure files
\usepackage{amsmath}    % Advanced maths commands
\usepackage{amssymb}    % Extra maths symbols
\usepackage{tabularx}    % For nice table
\usepackage{arydshln}    % For dashed line in table

\newcommand\lsim{\mathrel{\rlap{\lower4pt\hbox{\hskip1pt$\sim$}}
        \raise1pt\hbox{$<$}}}
\newcommand\gsim{\mathrel{\rlap{\lower4pt\hbox{\hskip1pt$\sim$}}
        \raise1pt\hbox{$>$}}}

\usepackage{xcolor}

\newcommand*{\colorboxed}{}
\def\colorboxed#1#{%
  \colorboxedAux{#1}%
}
\newcommand*{\colorboxedAux}[3]{%
  \begingroup
    \colorlet{cb@saved}{.}%
    \color#1{#2}%
    \boxed{%
      \color{cb@saved}%
      #3%
    }%
  \endgroup
}

\title[Combining UV LFs]{Combining high-$z$ galaxy luminosity functions with Bayesian evidence}

\author[N. J. F. Gillet et al.]{
Nicolas J. F. Gillet,\thanks{E-mail: nicolas.gillet@sns.it}
Andrei Mesinger 
and Jaehong Park.
\\
Scuola Normale Superiore, Piazza dei Cavalieri 7, I-56126 Pisa, Italy\\
}

\date{Accepted XXX. Received YYY; in original form ZZZ}

\pubyear{2018}

\begin{document}
\label{firstpage}
\pagerange{\pageref{firstpage}--\pageref{lastpage}}
\maketitle

\begin{abstract}
Galaxy formation during the first billion years of our Universe remains a challenging problem at the forefront of astrophysical cosmology.  Although these $z\gsim6$ galaxies are likely responsible for the last major phase change of our Universe, the epoch of reionization (EoR), detailed studies are possible only for relatively rare, bright objects. Characterizing the fainter galaxies which are more representative of the population as a whole is currently done mainly through their non-ionizing UV luminosity function (LF).  Observing the faint end of the UV LFs is nevertheless challenging, and current estimates can differ by orders of magnitude.  Here we propose a methodology to combine disparate high-$z$ UV LF estimates in a Bayesian framework: Bayesian Data-analysis Averaging (BDA).  Using a flexible, physically-motivated galaxy model, we compute the relative evidence of various $z=6$ UV LFs within the magnitude range $-20 \leq M_{\rm UV} \leq -15$ which is common to the data sets. Our model, based primarily on power-law scalings of the halo mass function, naturally penalizes systematically jagged points as well as misestimated errors.
We then use the relative evidence to weigh the posteriors obtained from disparate LF data sets during the EoR, $6 \leq z \leq 10$.
The resulting LF posteriors suggest that the star formation rate density (SFRD) integrated down to a UV magnitude of -17 represent $60.9^{+11.3}_{-9.6}\%$ / $28.2^{+9.3}_{-10.1}\%$ / $5.7^{+4.5}_{-4.7}\%$ of the total SFRD at redshifts 6 / 10 / 15. The BDA framework we introduce enables galaxy models to leverage multiple, analogous  LF estimates when constraining their free parameters.
\end{abstract}

\begin{keywords}
galaxies: high-redshift - reionization - first stars - early Universe 
\end{keywords}

%%%%%%%%%%%%%%%%%%%%%%%%%%%%%%%%%%%%%%%%%%%%%%%%%%

%%%%%%%%%%%%%%%%% BODY OF PAPER %%%%%%%%%%%%%%%%%%

%%%%%%%%%%%%%%%%%%%%%%%%
\section{Introduction}
\label{Sec:Introduction}
%%%%%%%%%%%%%%%%%%%%%%%%

The first billion years of the Universe remain a compelling cosmological mystery, mostly due to the fact that observations of this period remain challenging (e.g. \citealt{BarkanaLoeb2007, LoebFurlanetto2013, MesingerBook2016, Dayal2018}).
One of the simplest and most powerful observations are the non-ionizing ($\sim$1500\AA\ rest-frame) ultra-violet luminosity functions (UV LFs).  These can be obtained with relatively straightforward broad-band photometric drop-out techniques \citep{Steidel1999} and are thus useful in constraining the abundance of galaxies too faint to be studied with spectroscopy.

Nevertheless, pushing the UV LFs towards the fainter galaxies which are the dominant population during the first billion years is quite difficult.  Lensing magnification has been shown to be a powerful tool for this purpose; however, the systematics quickly become significant going towards magnification factors of beyond $\mu \gsim 10$ (e.g. \citealt{Bouwens17, Atek18}).  Various observational estimates of the faint end of the LF (where the bulk of the galaxies lie) can disagree by orders of magnitude.

How do we choose which LF estimates to use when constraining galaxy formation models?
If every competing LF estimate is analyzed independently, the corresponding parameter constraints would need to be eventually combined somehow.  Alternately, one could first combine different LF data sets in some fashion and then fit galaxy parameters to the {\it combined} data. For example, \citet{Finkelstein2016} used a Schechter function form \citep{Schechter1976} to fit LF estimates from various studies independently at each redshift ($z=$ 4--10).  These were then combined "agnostically": with each competing estimate being given the same weight, regardless of how consistent it was with the assumed Schechter form.

In principle, one should be able to improve on this by applying some basic, prior knowledge of what the UV LFs {\it should} look like.  
For example, sharp discontinuities in the LF would be very difficult to explain physically and could be an indication of an unaccounted for systematic in the observations.  The commonly used, empirically-motivated Schechter function requires additional, ad-hoc parameters to account for both redshift evolution and a faint-end turnover, in order to be consistent with observations and physically-motivated galaxy formation models of ultra-faint dwarf galaxies
(e.g. \citealt{Behroozi2013, Dayal2014, OShea2015, Liu2016, Finlator2017}).
%original citation list : \citealt{Jaacks2013, Behroozi2013, Paardekooper2013, Paardekooper2015, Dayal2014, OShea2015, Yue16, Liu2016, Gnedin2016, Ocvirk2016, Ocvirk2018, Wilkins2017, Finlator2017, Finlator2018, Cowley2018, Tacchella2018, Rosdahl2018, Ma2018, Ma2019, Yung19}).
Indeed, \cite{Yue18} use a physical galaxy model to supplement a Schechter function, and derive constraints on the presence of a faint-end turn-over in the LF, based on galaxy number counts at $z=6$.

{\it Here we use a flexible galaxy model to combine disparate high-$z$ LF estimates in a Bayesian evidence-based framework}.  The parametrization of this model should encapsulate the general, physical trends we expect from high-$z$ LFs, while still being able to accommodate the unknown details of galaxy formation.  We apply this Bayesian Data-analysis Averaging (BDA) framework to current LF data sets, resulting in combined LF constraints even at redshifts and magnitudes not probed by current observations.  

We note that the differences between current LFs mostly arise from different analysis methods applied to the same {\it Hubble} fields.  For example, different groups use different lensing models, completeness corrections, Eddington bias corrections, etc.  Therefore, LFs are not raw data, but derived data products. We therefore adopt the nomenclature Bayesian {\it Data-analysis} Averaging.

This paper is organized as follows. In \S \ref{Sec:Combining different observations} we describe the Bayesian Data-analysis Average method, demonstrating its use on toy LFs. In \S \ref{Sec:The non ionizing Luminosity function at high redshift}  we introduce the LF data sets and the analytic model used to discriminate between data sets.  In \S \ref{Sec:Results} we apply BDA on the $z=6$ LFs, and we use the resulting weights to combine LF data across $z\sim6$--10, presenting the resulting ``concordance'' LFs. In \S \ref{Sec:Conclusions} we state our conclusions.  Unless stated otherwise, we use comoving units, and assume the following $\Lambda$CDM cosmological parameters ($\Omega_m=0.3175$, $\Omega_{\Lambda}=0.6825$,
$h=0.6711$, $\Omega_b=0.049$, $n_s=0.9677$ and $\sigma_8=0.83$), consistent with the latest results from the {\it Planck} satellite \citep{Planckcosmo2018}, and magnitudes are given in the AB system. 

%%%%%%%%%%%%%%%%%%%%%%%%%%%%%%%%%%%%%%%%%%
\section{Combining different data sets}
\label{Sec:Combining different observations}
%%%%%%%%%%%%%%%%%%%%%%%%%%%%%%%%%%%%%%%%%%

Our methodology to combine the estimated LFs is inspired by Bayesian Model Averaging (BMA; e.g. \citealt{Parkinson2013}); however, we reverse ``model'' and ``data''.  Instead of comparing different models using a given observable, we compare different observables using a given model.  This comparison is done with the Bayesian evidence, which allows us to weigh the relative posteriors from different data sets and combine them using this weight.  We describe the procedure in detail below.

We note that alternative Bayesian methods have been proposed to combine data sets, taking advantage of Bayesian hierarchical modeling and/or hyper-parameters.
A common approach is to add hyper-parameters to account for mis-estimated errors / systematics of each observation, which are then marginalized over to obtain the posterior of the desired quantities (e.g. \citealt{Lahav2000, Hobson2002, MA2014, Bernal2018}).
Such an approach relies on knowing how to parametrize these uncertainties and the additional parameters make the likelihood calculation more expensive. The procedure we propose
avoids this but at the cost of relying on a parametrization of the "truth".

Below we briefly review BMA, before introducing our reversed application of it: BDA.  We then demonstrate its use using toy models for LFs.

%%%%%%%%%%%%%%%%%%%%%%%%%%%%%%%%%%%%
\subsection{Bayes' equation and model averaging}
\label{Sec:Bayes' equation and model averaging}
%%%%%%%%%%%%%%%%%%%%%%%%%%%%%%%%%%%%

Let $\mathcal{D}$ be a data catalog composed of several data sets (raw or derived observations) and $\mathcal{M}$ an analytic model with parameters $\rm{\theta}$.
       Bayes' theorem permits us to compute the posterior: the probability distribution of the parameters $\rm{\theta}$ given a specific data set $\mathcal{D}_i$:
    \begin{equation}
        \rm{ P(\theta | \mathcal{D}_i, \mathcal{M}) = \frac{P(\mathcal{D}_i | \theta, \mathcal{M}) P(\theta | \mathcal{M})}{P(\mathcal{D}_i |  \mathcal{M}) = \int_{\theta} P(\mathcal{D}_i | \theta, \mathcal{M}) P(\theta | \mathcal{M}) d\theta } },
    \label{Eq:BayesTheo}
    \end{equation}
    where $\rm{ P(\theta | \mathcal{M}) }$ is the prior on the parameters, $\rm{ P(\mathcal{D}_i | \theta, \mathcal{M}) }$ is the likelihood (commonly based on $\chi^2$), and $\rm{ P(\mathcal{D}_i |  \mathcal{M}}$) is probability of the data given the model (also known as the evidence).
    
      In general, the posterior is just the normalized likelihood distribution, weighted by the priors. The evidence is commonly used only as a normalization factor because one is interested in the relative probabilities across the parameter space of $\theta$.  However, if one has various competing models,  $\mathcal{M}_i$, then the relative evidence can be used to discriminate among them, answering the question: ``which model is preferred by the data?''.  Additionally, the evidence can be used to average over parameters common to the various models.  This is referred to as Bayesian model averaging (BMA).
    
%%%%%%%%%%%%%%%%%%%%%%%%%%%%%%%%%%%%
\subsection{Bayesian Data-analysis Averaging}
\label{Sec:Bayesian Data-analysis Averaging}
%%%%%%%%%%%%%%%%%%%%%%%%%%%%%%%%%%%%

In this work, we invert "data" and "model", asking the question: ``which data set is preferred by our model?''    Given a model $\rm{ \mathcal{M} }$, we can compute the relative evidence of the data sets: 
    \begin{equation}
        \rm{  P(\mathcal{D}_i | \mathcal{D}, \mathcal{M} ) = \frac{P(\mathcal{D}_i| \mathcal{M})}{ \sum_j P(\mathcal{D}_j |  \mathcal{M})} },
    \label{Eq:relative_evidence}
    \end{equation}
    
   The prior on the data set $\pi{(\mathcal{D}_i) }$ is taken to be uniform.
    This relative evidence can be used to compare the data sets between each other, given the model.  We use the relative evidence from each data set as a weight of the resulting posterior for our model parameters:
    \begin{equation}
        \rm{  P(\theta | \mathcal{D}, \mathcal{M}) = \sum_i P( \theta | \mathcal{D}_i, \mathcal{M} ) \times P(\mathcal{D}_i | \mathcal{D}, \mathcal{M} ) },
    \label{Eq:BDA_combining}
    \end{equation}
    Where $\rm{  P(\theta | \mathcal{D}, \mathcal{M} )}$ is the final constrained posterior distribution.  The corresponding ``concordance'' LF is then obtained by sampling this combined posterior.   

    It is important to keep in mind that this procedure is model dependent.  Ideally, one should choose a model with a parametrization capable of capturing the general trends we expect from the data, yet flexible enough to accommodate the large range of uncertainties.  Conceptually, this is analogous to putting a (conservative) prior on what is expected from the observations.  The model we use for this purpose is described in \S \ref{Sec:Analytic model}.

%%%%%%%%%%%%%%%%%%%%%%%%%%%%%%%%%%%%%%%%%%%%%%%%%%%%%%%
\subsection{Demonstration on toy models}
\label{Sec:Demonstration on toy models}
%%%%%%%%%%%%%%%%%%%%%%%%%%%%%%%%%%%%%%%%%%%%%%%%%%%%%%%

\begin{figure*}
\includegraphics[width=\columnwidth]{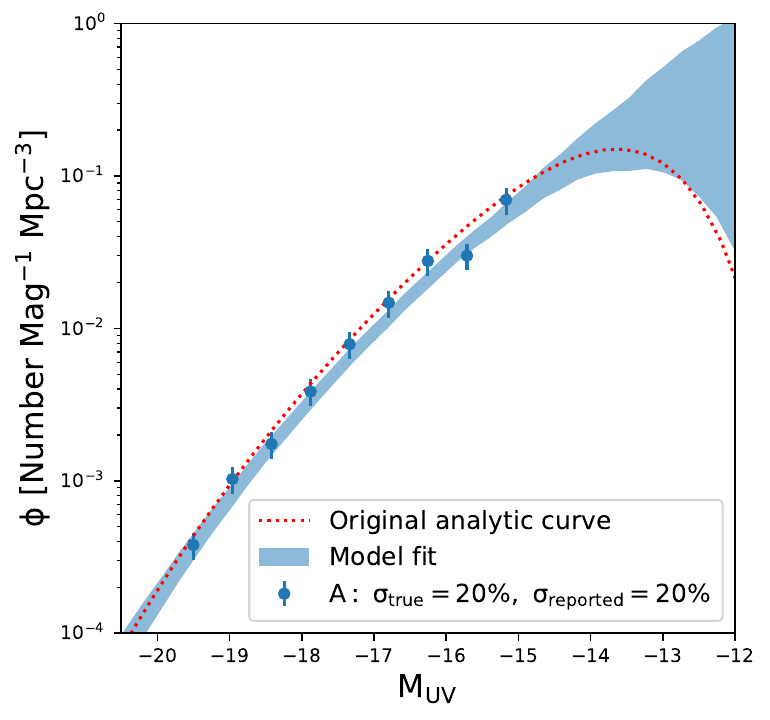}
\includegraphics[width=\columnwidth]{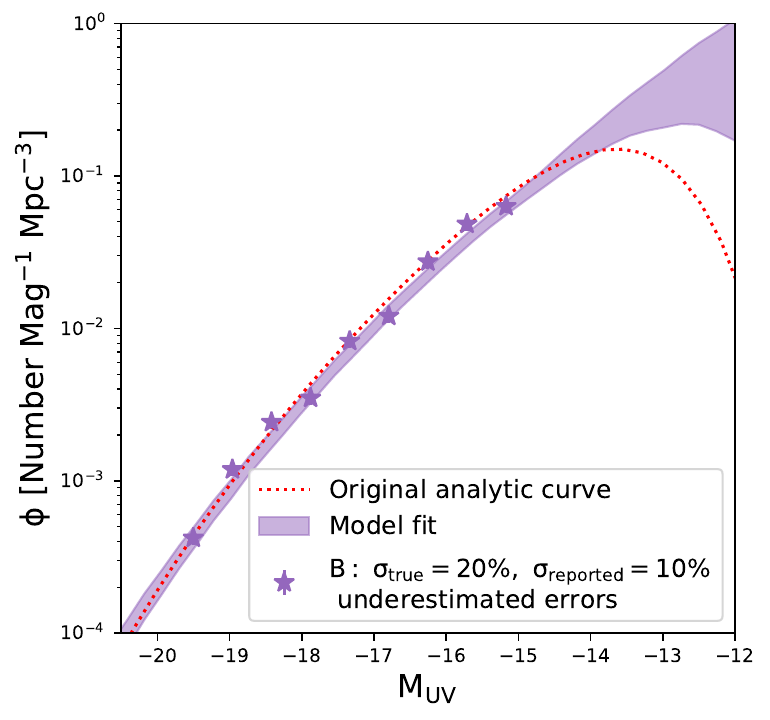} \\ 
\includegraphics[width=\columnwidth]{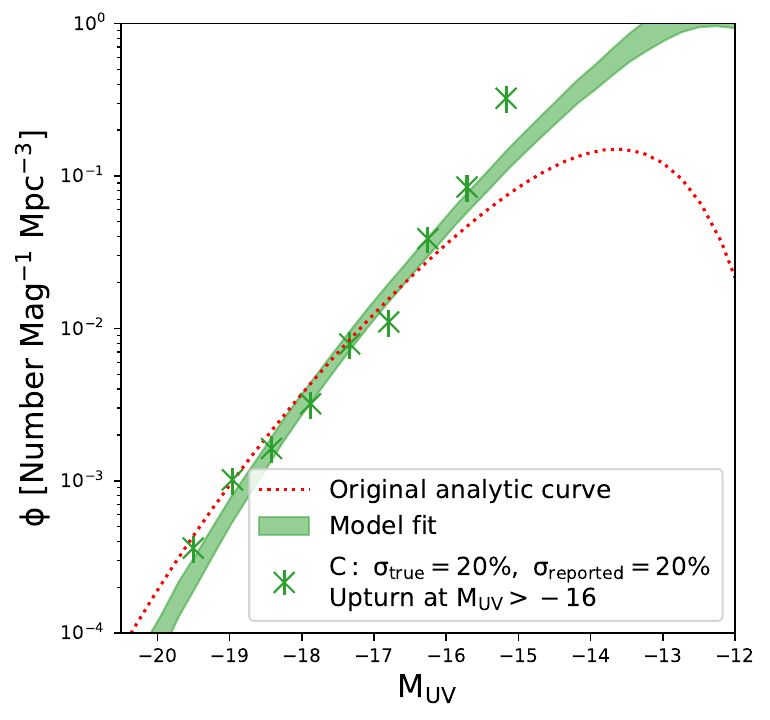}
\includegraphics[width=\columnwidth]{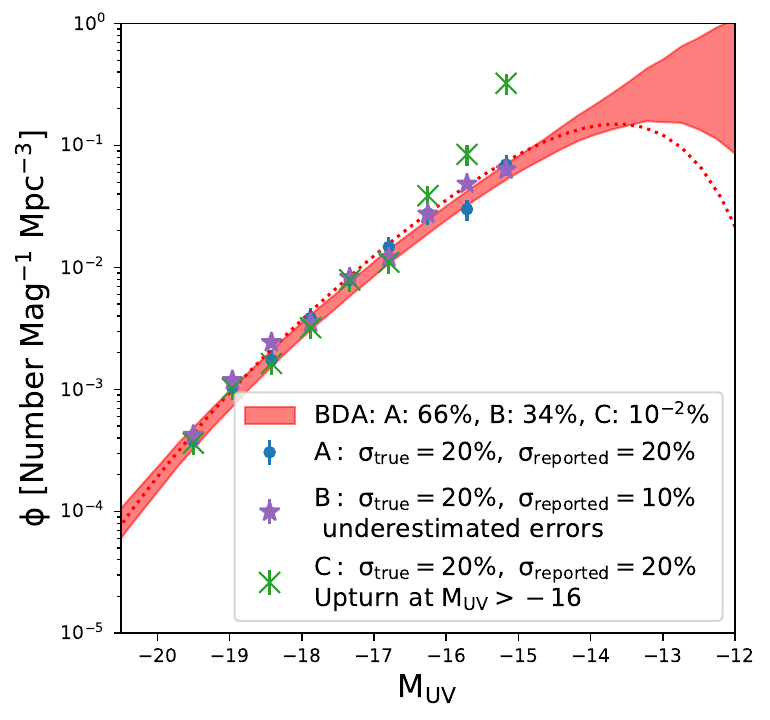}
\caption{ Mock luminosity functions to demonstrate the application of Bayesian Data-analysis Averaging. The three mock data sets correspond to the dots (stars and X) with the reported errors. The red dashed line illustrates the true underlying LF used to create the mocks. The shaded areas represent the 68\% confidence interval from the posterior.  On the bottom right, the same three mocks are shown with colored dots (stars and X) and the shaded area corresponds to the BDA combined posterior, with the relative weights of the three data sets shown in the legend.}
\label{Fig:toy_model}
\end{figure*}

 Here we illustrate the use of BDA, applied on toy LFs.  Our mock LFs consist of nine points, generated by different methods of sampling a fiducial parameter combination (\S \ref{Sec:Analytic model}):
    \begin{itemize}
    \item Mock observation (A) was generated by sampling the expectation values from this model, assuming Gaussian errors with a standard deviation of 20\%, for each magnitude bin.  The reported errors on these points also have a standard deviation of 20\%.  Thus, the samples are consistent with the underlying model and the reported uncertainty corresponds to the true uncertainty.  Hence, model (A) represents an accurate data set (c.f. top left panel in Fig. \ref{Fig:toy_model}).
\end{itemize}
\begin{itemize}
    \item Mock observation (B) was generated by sampling the same analytic model as (A), also taking Gaussian errors with a standard deviation of 20\%.  However, here the reported errors are underestimated to be only 10\% (c.f. top right panel in Fig. \ref{Fig:toy_model}).
    \end{itemize}
    \begin{itemize} 
    \item Mock observation (C) is statistically the same as (A) for the brightest six points; however, the faintest three data points are systematically offset from the underlying analytic model, showing an upturn for $M_{\rm{UV}}>-16$ of 15\%, 30\%, and 50\%, respectively.  This observation is illustrative of an unknown systematic in the data, which cannot be captured by our model  (c.f. bottom left panel in Fig. \ref{Fig:toy_model})
    \end{itemize}
    
    We show the three mock data sets and 68\% confidence interval (C.I.) on the posteriors in the first three panels of Fig. \ref{Fig:toy_model}.  As expected, the posteriors of data set A and B are comparable, given that they only differ in the error estimates.  However model C prefers a much steeper LF posterior.  This is because our galaxy model (described in detail in \S 3.2) does not allow for upturns\footnote{
    \cite{Mirocha2019} use an empirical model to show that {\it if} the recent EDGES claimed detection \citep{Bowman2018} was cosmological, then one would need either an upturn in the LFs or a new population of faint, transient galaxies at $z\sim20$. Allowing for an upturn in LFs could alter some conclusions here, and we defer it to future work.  We point out however, that an upturn at $z\sim6$ is very contrived.  The most plausible physical motivation of such an upturn could be ultra faint dwarfs, which get their gas through molecular hydrogen cooling (so-called minihalos). These form from pristine gas and could have different properties from the observed galaxies  (e.g. \cite{Tumlinson2000, Schaerer2002, Yoshida2008, Xu2016b, Koh2018}), which could result in an upturn. However, these minihalos likely lie at fainter magnitudes than those accessible with current observations (e.g. \cite{O'Shea2015} and Qin et al. in prep).  Moreover, minihalos are expected to disappear well before $z~6$, since they get sterilized quickly by a Lyman Werner background ( \cite{Holzbauer2012, Fialkov2013} and Mebane et al. in prep).}
    , and so the last three points steepen the LF posterior, despite the fact that the first 6 points are statistically the same as for model A.

    In the final panel, we show the combined LF posteriors, obtained after using the relative evidence to weight the posteriors of A, B and C (eq. \ref{Eq:BDA_combining}).  The relative evidence from BDA is shown in the legend:  66\%, 34\%, $\rm{ 10^{-2} }$\%, for data sets A, B and C respectively.  BDA down-weighs the posterior of data set C quite strongly, and so it does not really contribute to the combined posterior.  This ``penalty'' is due to our {\it belief} (qualified in terms of our analytic model; see Section \ref{Sec:Analytic model}), that upturns in LFs are nonphysical.

    Data set A provides the most constraining power, as the error bars of the data are estimated properly.  BDA prefers A over B by a factor of two, even though the only difference between the two data sets is that the later data set underestimated the errors of its data points.\footnote{We repeat this experiment with 1000 different realizations, finding that data set A consistently contributes the most to the combined posterior, at the level of 70\% on average.}

%%%%%%%%%%%%%%%%%%%%%%%%%%%%%%%%%%%%%%%%%%%%%%%%%%%%%%%%%%%%%%%
\section{The non-ionizing Luminosity function at high redshift}
\label{Sec:The non ionizing Luminosity function at high redshift}
%%%%%%%%%%%%%%%%%%%%%%%%%%%%%%%%%%%%%%%%%%%%%%%%%%%%%%%%%%%%%%%

We now wish to apply BDA on actual LF estimatess. We first discuss the data sets we use, then our analytic model which is used to weigh them, before specifying how we compute the evidence.

%%%%%%%%%%%%%%%%%%%%%%%%%
\subsection{Luminosity Function estimation}
\label{Sec:Observations}
%%%%%%%%%%%%%%%%%%%%%%%%%

\begin{figure*}
\includegraphics[width=\textwidth]{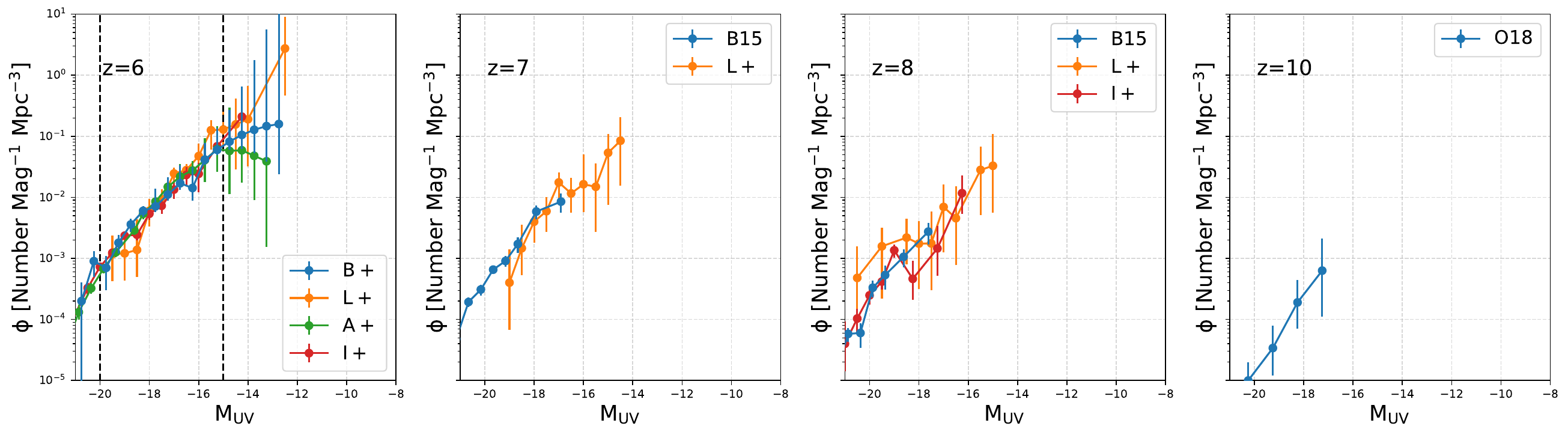}
\caption{ LF determinations at redshifts 6, 7, 8 and 10 from left to right (see text for details and references). The vertical black dash lines delimit the "faint" and "ultra-faint" end. }
\label{Fig:frise_OBS}
\end{figure*}

\begin{figure}
\includegraphics[width=\columnwidth]{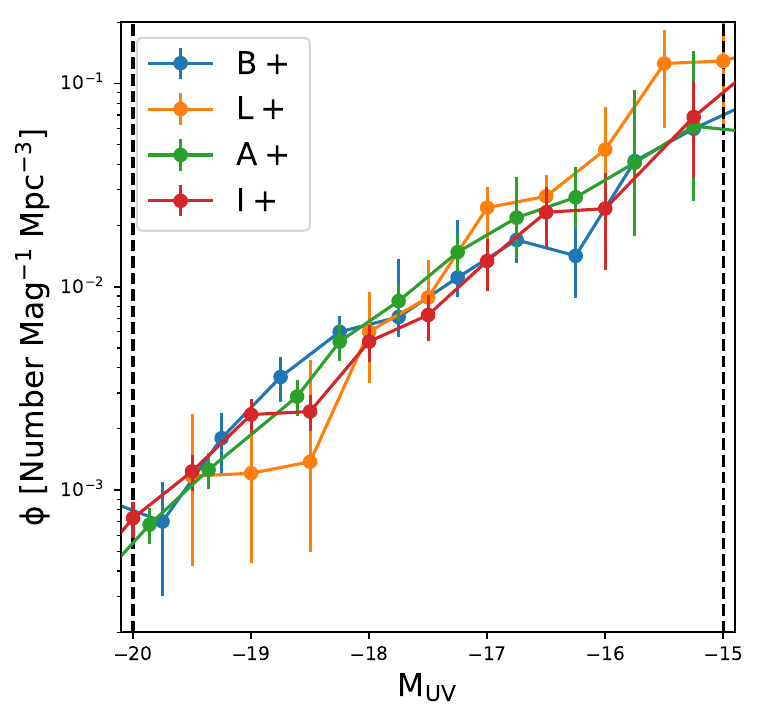}
\caption{ Zoom-in on the ten $z=6$ LFs points that are common to all data-sets, and which we use when computing the BDA evidence. }
\label{Fig:OBS_zoom}
\end{figure}

In this study, four sets of high-$z$ LFs are used, from redshift 6 and above when available. In the rest of the paper we define "faint end" to be magnitudes fainter than $\rm{ -20 }$ (the dominant population we are interested in characterizing) and "ultra-faint end" to be magnitude fainter than $\rm{ -15 }$ for which lensing uncertainties increase dramatically (c.f. \citealt{Finkelstein2016, Bouwens17, Atek18}). These four data sets are:

\begin{itemize}
    \item The \underline{"Bouwens et al. data set" (B+)}: consisting of the $z=6$ LF from \cite{Bouwens17}, the $z=7$ and $8$ LFs from \cite{Bouwens15}, and the $z=10$ LF from \cite{Oesch18}. The estimates at $z=6$ are based on the four first clusters of the Hubble Frontier Field program (HFF): Abell 2744, MACS0416, MACS0717, and MACS1149.
    
    \item The \underline{"Atek et al. data set" (A+)}: we take the reported LF from \cite{Atek18}, adjusted according to their prescription to correspond to $z=6$. This data set used the six clusters of HFF: Abell 2744, MACS0416, MACS0717, MACS1149, AS1063 and A370 and in addition they use the bright part of the LF from \cite{Bouwens15}.
    
    \item The \underline{"Ishigaki et al. data set" (I+)}: consisting of the $z=6$ and 8 LFs from \cite{Ishigaki18}. They use the four first HFF clusters, as well as the LF, extracted from blank fields from \cite{Bouwens15}.
    
    \item The \underline{"Livermore et al. data set" (L+}): consisting of the $z=6$, 7, 8 LFs from Livermore (private communication; Finkelstein in prep).  The LFs correspond to those in \cite{Livermore17}, but corrected for Eddington bias, which reduces the implied number densities, most notably at the faint end by up to a factor of $\sim$ 2 (though we note that the amplitude of Eddington bias corrections is debatable, with \cite{Bouwens17} arguing they should be no larger than 10\%).  These Eddington-bias adjusted LFs have also been used in \cite{Yung19}. The two first HFF clusters are used to derive the faint end LF: Abell 2744 and MACS0416.
\end{itemize}

We assume a minimum fractional uncertainty of 20\% (in linear scale), as suggested in \cite{Bouwens17}, increasing the error of all the data points if the reported error is smaller. 
Figure \ref{Fig:frise_OBS} presents these four data sets, at redshift 6, 7, 8 and 10 from left to right.  As seen in the panels, the implied galaxy density can vary by orders of magnitude, especially in the ultra-faint end when lensing uncertainties such as completeness corrections dominate the systematics.

To compute the relative evidence as described above, we need data at the same magnitude and redshift bins.  For this purpose, we use the ten points in the magnitude range $-20 \leq M_{\rm UV} \leq -15$ at $z=6$ (c.f. Fig. \ref{Fig:OBS_zoom}). The bright limit of this range is still faint enough to be relatively free from dust and AGN feedback, which are not accounted for in our model.  Indeed the slope of the UV continuum $\beta$ seems to change around this value above redshift 6 (e.g. \citealt{Finkelstein12, Bouwens2014}), roughly consistent with simulation results which suggest that at fainter magnitudes the impact of dust starts becoming negligible (e.g. \citealt{Cullen2017, Wilkins2016, Wilkins2017, Ma2019}, and AGN feedback can be neglected (e.g. \citealt{Wilkins2017, Yung19})\footnote{ We test the impact of the bright end limit on our results by removing the brightest two magnitude bins and re-computing the posteriors.  The resulting posteriors are consistent with our fiducial ones, with a somewhat broader PDF for the slope parameter, $\alpha_*$, due to the removal of points with comparably small errors.  Thus we do not find evidence that the bright end limit changes the implied slope of the stellar mass to halo mass relation, and as a consequence that we would need additional parameters characterizing dust or AGN feedback.  The relative evidence does change somewhat for this reduced data set, with 19\% / 37\% / 41.5\% / 2.5\% attributed to B+ / I+ / A+ / L+  (c.f. values in Table 1).  This reflects the fact that the I+ data set has very small errors for those two bins, and the implies counts are consistent with our parametrization. Thus their removal shifts some of the corresponding relative evidence to B+.  Selecting sub-samples of the data is, in any case, ad-hoc, so we use the largest range which is common to the data sets and over which our galaxy parametrization is reasonable.}.
The faint limit, although in the lensing regime ($M_{\rm UV} \gsim -17$) is sourced by relatively modest magnification factors, with correspondingly well-behaved uncertainties \citep{Finkelstein2016, Bouwens17, Atek18}.  Most importantly, this range is common to all four data sets, which is necessary in order to compare their corresponding Bayesian evidence.

Unfortunately, by calculating the evidence over a limited range, we end up implicitly assuming that this range characterizes also the systematic biases in the other magnitude bins used when combining the data sets.  This is unlikely to be true; however our fiducial choice of $-20 \leq M_{\rm UV} \leq -15$ at $z=6$ corresponds to the maximum range all data sets have in common.  In the absence of a compelling a-priory reason to perform further data cuts, it is reasonable to use the largest range possible (see however the test in footnote 3).  Moreover, we note that these data are the most constraining, as the fainter magnitudes and higher redshift data points contribute only modestly to the posterior. As discussed in detail in Appendix \ref{App:Comparison BDA and Average} and shown in Fig. \ref{Fig:Comparison BDA AVG}, the addition of fainter magnitudes mainly results in a slight peak in the posterior for the turn-over scale and as a small bi-modality in the star-formation rate to halo mass scaling, both driven by the A+ data set.

%%%%%%%%%%%%%%%%%%%%%%%%%%%
\subsection{Analytic model}
\label{Sec:Analytic model}
%%%%%%%%%%%%%%%%%%%%%%%%%%%

\begin{figure}
\includegraphics[width=\columnwidth]{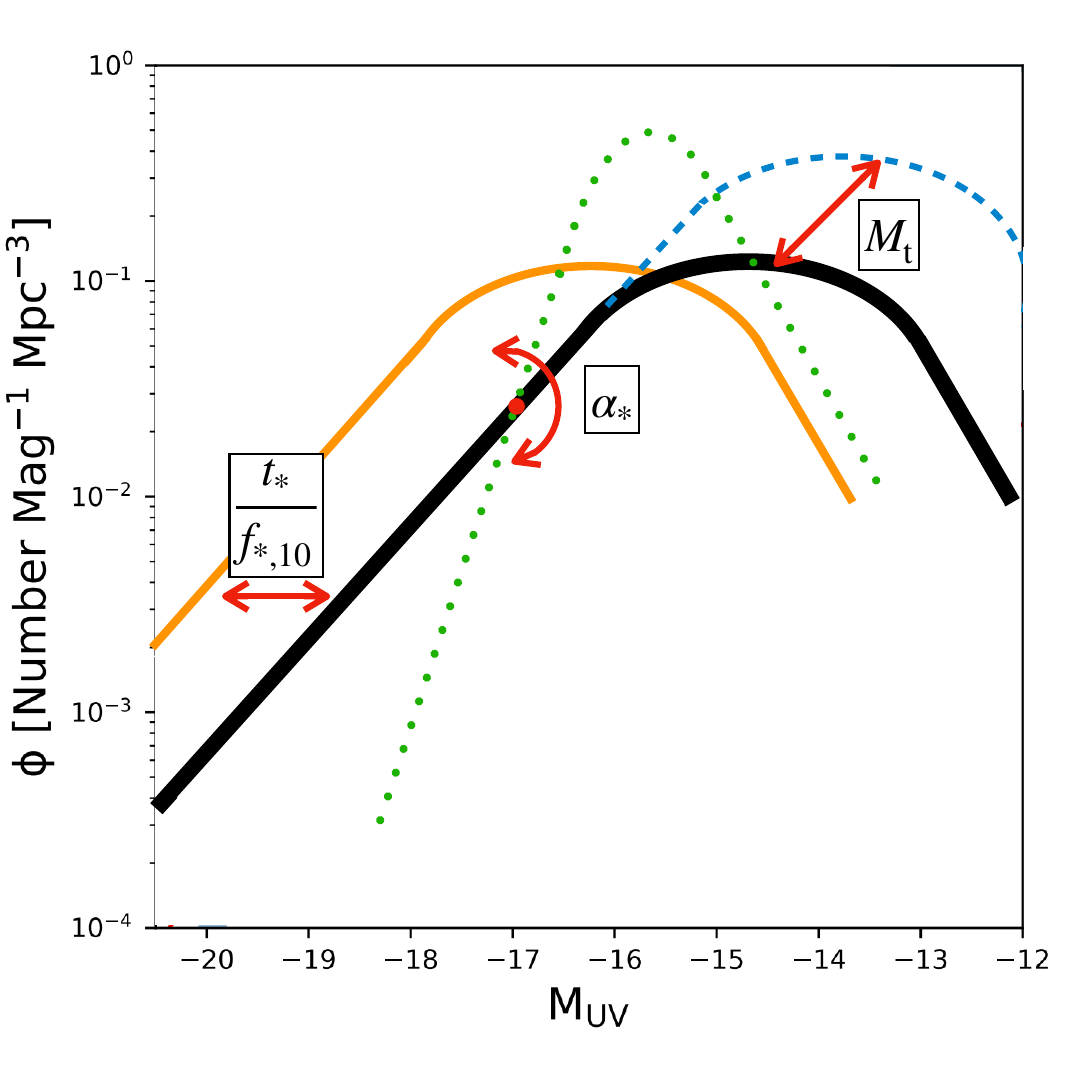}
\caption{ Schematic view of the influence of the model's parameters on the LF. See the discussion in the text and Park et al. 2019 for more details. }
\label{Fig:LF_scheme}
\end{figure}

The analytic model, $\mathcal{M}$, used in this study is the same as in \cite{Park18}.  This model characterizes UV LFs using four, fairly empirical parameters.  It is physically motivated in the sense that it scales the LF from the halo mass function (HMF), assuming power-law scalings.  Specifically, the typical stellar mass, $M_\ast$, of galaxies residing in halos of total mass, $M_{\rm h}$, is assumed to (on average) follow a power-law with arbitrary amplitude and power law index (see \citealt{Behroozi2013, Behroozi2015}) :
\begin{equation}
    M_\ast(M_{\rm{h}}) = f_{*,10} \left( \frac{M_{\rm{h}}}{10^{10}\rm{M_{\odot}}} \right)^{\alpha_*}  \left( \frac{ \Omega_{\rm{b}} } { \Omega_{\rm{m}} } \right) M_{\rm{h}}.
\end{equation}

The typical star formation rate (SFR) in a given halo mass bin is taken to be the total stellar mass divided by some fraction of the Hubble time:
\begin{equation}
    \dot{M}_*(M_{\rm{h}},z) = \frac{M_*}{t_* H^{-1}(z)},
\end{equation}

The SFR is then converted to a UV luminosity assuming a simple conversion factor:
\begin{equation}
    \dot{M_*}= \kappa_{\rm{UV}} \times L_{\rm{UV}}, 
\label{Eq:SFR_lum}
\end{equation}
where $\kappa_{\rm{UV}}=1.15\times 10^{-28} \rm{M_{\odot}yr^{-1} / erg s^{-1} Hz^{-1}}$ (\citealt{Sun17}; see also \citealt{Kennicutt1998, Madau2014, Bouwens2012}) is determined by the IMF (and is degenerate with our SFR parameters) and the UV magnitude is computed from the UV luminosity: 
\begin{equation}
    {\rm{log_{10}}} (L_{\rm{UV}}) = 0.4 \times ( 51.63 - M_{\rm{UV}} ).
\end{equation}

Star formation in low mass halos is suppressed via a  "duty cycle", motivated by inefficient gas accretion and/or strong feedback (e.g. \citealt{Okamoto2008, Sobacchi2013, Sobacchi2014, Dayal2014, OShea2015, Yue16, Ocvirk2016, Ocvirk2018}).
Specifically, we assume that only a fraction $f_{\rm duty}$ of halos of mass $M_{\rm h}$ can host star-forming galaxies, with:
\begin{equation}
    f_{\rm{duty}} (M_{\rm{h}}) = {\rm{exp}} \left( - \frac{ M_{\rm{t}} }{ M_{\rm{h}} } \right).
\end{equation}
Here, $M_{\rm{t}}$ is the characteristic halo mass scale below which star formation is inefficient.  Our results are not very sensitive to the exact functional form of this duty cycle, since most of the observations probe galaxies inside more massive halos, as we shall see below.

Finally, the LF is computed from the halo mass function and the relation between the halo mass and the UV magnitude:
\begin{equation}
    \phi(M_{\rm{UV}}) = \left( f_{\rm{duty}}\frac{ {\rm{d}} n }{ {\rm{d}} M_{\rm{h}} }  \right) \frac{ {\rm{d}} M_{\rm{h}} }{ {\rm{d}} M_{\rm{UV}} } .
\end{equation}

The model has 4 free parameters: $f_{\rm{*,10}}$, $\alpha_*$, $t_* $ and $M_{\rm{t}}$. Hereafter we call a parameter point: $\theta = (f_{\rm{*,10}},\ \alpha_*,\ t_*,\ M_{\rm{t}} )$.  Figure \ref{Fig:LF_scheme} presents a schematic view of the influence of each parameter on the LF. \footnote{An animation showing how the parameters influence the LFs and  the 21-cm signal is available at \url{http://homepage.sns.it/mesinger/Videos/parameter\_variation.mp4}.} 
The turn-over mass $M_{\rm{t}}$ shifts the peak of the LF towards fainter and brighter magnitudes.  The parameter $\alpha_*$ controls the slope, rotating the LFs around the normalization value of $M_{\rm{UV}}(M_{\rm{h}}=10^{10}M_{\odot})$, which depends on $f_{\rm{*,10}}$ and $t_*$.  It also changes the location of the turn-over, because it affects the conversion of halo mass to magnitude (note that the model parameters are all defined in terms of halo mass, not directly a UV magnitude).  Finally, the parameters $f_{\rm{*,10}}$ and $t_*$ translate the LF curves horizontally.  Note that these two parameters are completely degenerate, and the LF is only sensitive to the ratio ${t_\ast}/{f_{\rm{*,10}}}$, though ancillary observations such at 21-cm can mitigate this degeneracy.  
We refer the reader to \cite{Park18} for a detailed analysis of the influence of each parameter on the luminosity function.

The important point for this study is that (i) this model is physically motivated: the galaxy density is directly linked to the dark matter halo density allowing us to penalize extreme LF shapes which are difficult to obtain from HMFs; and (ii) the model is flexible enough to fit reasonably well a large variety of estimated luminosity functions as well as those from hydrodynamic cosmological simulations (see Appendix 1 in \citealt{Park18}) and SAMs (Greig et al., in prep).

%%%%%%%%%%%%%%%%%%%%%%%%%%%%%%%%%%%%
\subsection{Computing the likelihood and the evidence}
\label{Sec:Computing the likelihood and the evidence}
%%%%%%%%%%%%%%%%%%%%%%%%%%%%%%%%%%%%

Computing the evidence can be computationally challenging in high-dimensional parameter space (e.g. \citealt{Trotta2008}), since the likelihood has to be integrated over the whole space (see the denominator of Eq. \ref{Eq:BayesTheo}).  

Our likelihood at each parameter sample $\theta$ is computed by comparing the corresponding (logarithmic) LF to the data at every magnitude and redshift bin considered.  We use a split-norm probability density function around each data point to take into account of the asymmetrical errors in most of the data-sets (c.f. Appendix \ref{App:Split Norm}). Explicitly, we have:

\begin{equation}
    P(\mathcal{D}_i | \theta ) = \prod_{M_{\rm{UV}}\ bins} \mathcal{S}(x(\theta),\mu,\sigma_1,\sigma_2),
\end{equation}

\noindent where $\mathcal{S}$ is the split-norm likelihood (see eq. A1), $x(\theta)$ is the modeled galaxy density in a given bin and $(\mu,\sigma_1,\sigma_2)$ the data galaxy density and its positive and negative errors in the same bin.

We have to compute the likelihood over the whole 4D parameter space $\theta = [f_{\rm{*,10}},\ \alpha_*,\ t_*,\ M_{\rm{t}} ]$.  Running an MCMC like in \cite{Park18} for a dataset takes 2 weeks.  In this study we want to compute the likelihood distribution for 4 datasets, and explore different configurations.  To aid in this computation, we pre-compute the $z=$ 6, 7 , 8 and 10  LFs corresponding to a grid of $4\times10^5$ parameter space samples.   This grid is the concatenation of 8 Latin Hyperbolic Samples (LHS) of 50.000 points each.  The construction of this pre-computed grid of LFs takes around 2 weeks, but the subsequent likelihood calculation is very fast: it just takes a dozen minutes to obtain the likelihood distribution over 400000 points for each data set on a single core, while an MCMC with the same chain length could take days on several cores.  We check that this approximation of the likelihood distribution is converged by comparing with the MCMC results from \cite{Park18}, (see Appendix \ref{App:Convergence test}). We also check that the posterior is unchanged when computed using only half of the grid samples (i.e. 200000 points).  The discreteness of the sampling results in some noticeable noisiness in the {\it marginalized} posteriors; however, the parameter estimation and the evidence is converged (see Appendix \ref{App:Convergence test}).

Finally, once the likelihood samples are computed, the evidences can be estimated.  We interpolate the 400000 likelihood samples to an evenly spaced grid, consisting of 50 bins per axis, and compute the evidence by summing over this grid.  We consider uniform priors for all the parameters over the ranges $( \rm{log_{10}}(f_{\rm{*,10}}) \in [-2.5,-1],\ \alpha_* \in [-0.5,1],\ t_* \in [0,1],\ \rm{log_{10}}(M_{\rm{t}}) \in [8,10] )$.

%%%%%%%%%%%%%%%%%%%
\section{Results}
\label{Sec:Results}
%%%%%%%%%%%%%%%%%%%

\renewcommand{\tabularxcolumn}[1]{>{\centering\arraybackslash}m{#1}}
\begin{table}
    \begin{center}
        \begin{tabularx}{\columnwidth}{XXXX}
            \hline
            (B+) & (I+) & (A+) & (L+) \\ 
            3.5   & 52.9  & 43.4  & 0.2  \\ 
            \hline
        \end{tabularx}
    \end{center}
\caption{ Relative evidence of the data sets in the magnitude range $\rm{[-20,-15]}$ at redshift 6, given the analytic model (in \%). }
\label{Tab:BDA_evidence}
\end{table}

\begin{figure*}
\includegraphics[width=0.8\textwidth]{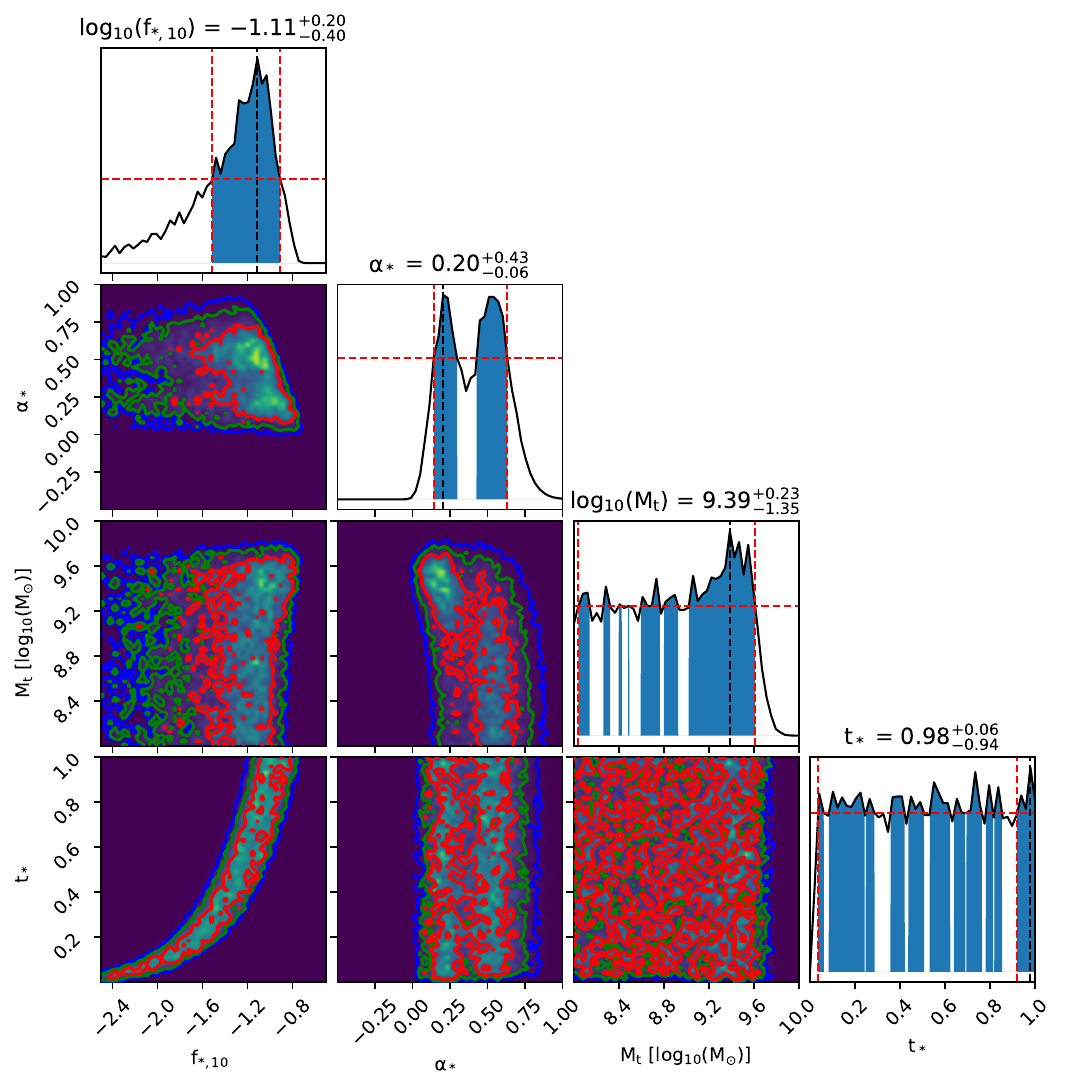}
\caption{ 1D and 2D marginalized posterior distributions of galaxy parameters resulting from the BDA weighing of the posteriors from each data set. The relative weights are listed in table \ref{Tab:BDA_evidence}.  Although the relative weights are computed using only the ten LF points at $z=6$ common to every data set (see Fig. \ref{Fig:OBS_zoom}), the final posteriors that are averaged are then re-computed using all data points (see text for details). The resulting distribution is mostly the average of the posteriors of A+ and I+. In the 1D marginalized figures, the reported values are the maximum and 68\% of highest posterior density (HPD) (illustrated by the blue shaded area).}
\label{Fig:posterior}
\end{figure*}

\begin{figure*}
\includegraphics[width=\textwidth]{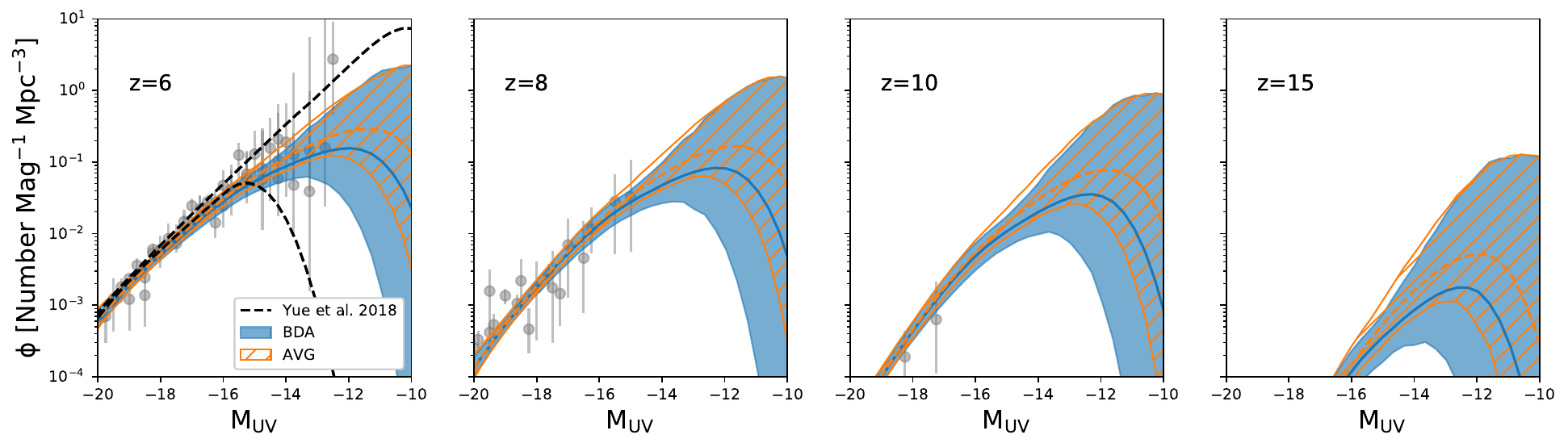}\
\includegraphics[width=\textwidth]{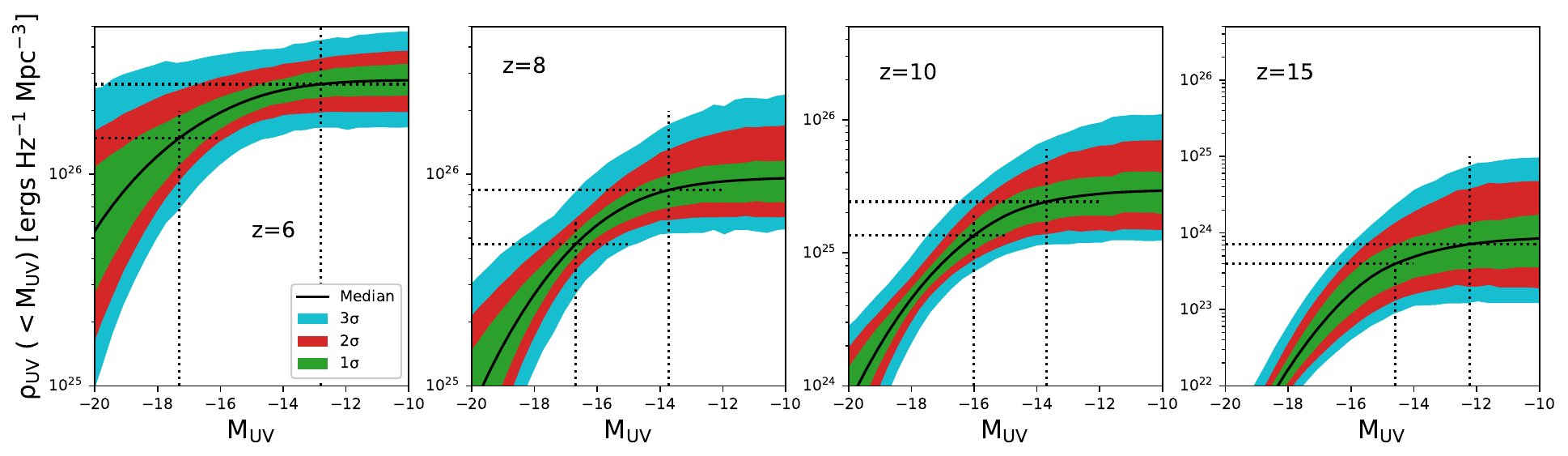}
\caption{ {\it First row}: the 68\% confidence interval of the combined LFs corresponding to the BDA posteriors from Fig. \ref{Fig:posterior} ({\it blue shaded regions}). For comparison, the orange shaded regions show LF constraints if instead of BDA weights (see table \ref{Tab:BDA_evidence}), the posterior of each data set were given an equal weight (i.e. an average of the posteriors) (orange hatched area).  In this later case, the relative down-weighting of A+ evidenced by the turnover scale shifting towards fainter magnitudes.  All data points used in this work are shown as the grey dots. In the first panel, the 68\% C.L. from the data-constrained model of \citealt{Yue18} are shown with the dashed black lines.} {\it Second row}: the 68\%, 95\% and 99\% confidence limits of the cumulative UV luminosity density corresponding to the BDA LFs. The dashed lines correspond to the magnitude limit below which brighter galaxies contribute  50\% and 90\% of the total UV luminosity density.
\label{Fig:BDA_frise}
\end{figure*}

\begin{figure}
\includegraphics[width=\columnwidth]{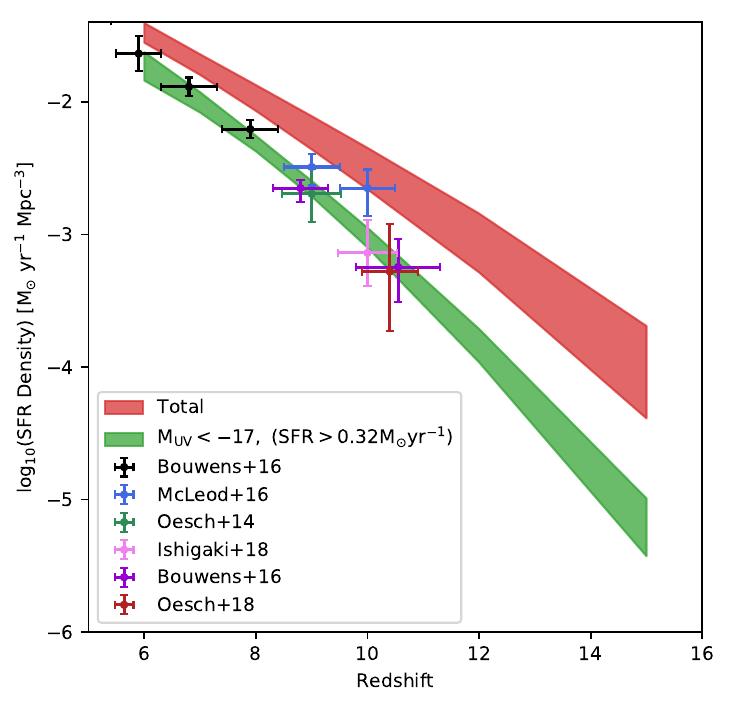}
\caption{ 68\% C.L. on  the cosmic SFR density implied by our BDA LFs, integrated down to $M_{\rm{UV}}<-17$ (corresponding to SFR $\gsim0.32\ \rm{M_{\odot}\ yr^{-1}}$ ({\it green shaded area}), as well as the total SFRD ({\it red shaded area}). The estimated data sets have been homogenized by considering the same SFR-luminosity relation (Eq. \ref{Eq:SFR_lum}) and the same integration limit of -17 (data provided by Oesch priv. com. and published in \citealt{Oesch18}).  The derived SFRD for the two thresholds are given in the table \ref{Tab:SFRD}, as well as the completeness. The original data points are from \citealt{Bouwens2014, Bouwens2016, McLeod2016, Oesch2013, Oesch2014, Ishigaki18}. }
\label{Fig:SFRD}
\end{figure}

\begingroup
\renewcommand{\tabularxcolumn}[1]{>{\centering\arraybackslash}m{#1}}
\renewcommand{\arraystretch}{1.5}
\begin{table*}
    \begin{center}
        \begin{tabularx}{\textwidth}{c|XXXXXXX}
            \hline
            Redshifts & 6 & 7 & 8 & 9 & 10 & 12 & 15 \\
            \hline
            SFRD at $ M_{\rm{UV}} <-17 $ & $-1.72^{+0.12}_{-0.10}$ & $-2.01^{+0.07}_{-0.08}$ & $-2.32^{+0.06}_{-0.06}$ & $-2.66^{+0.06}_{-0.06}$ & $-3.02^{+0.08}_{-0.07}$ & $-3.82^{+0.14}_{-0.11}$ & $-5.20^{+0.23}_{-0.21}$  \\
            SFRD total               & $-1.49^{+0.07}_{-0.08}$ & $-1.71^{+0.08}_{-0.07}$ & $-1.96^{+0.11}_{-0.08}$ & $-2.21^{+0.15}_{-0.10}$ & $-2.47^{+0.19}_{-0.13}$ & $-3.04^{+0.24}_{-0.20}$ & $-4.00^{+0.38}_{-0.31}$  \\
            SFRD completeness at $ M_{\rm{UV}} <-17 $ (in \%) & $60.9^{+11.3}_{-9.6}$ & $52.6^{+11.0}_{-9.3}$ & $44.1^{+10.6}_{-9.2}$ & $36.0^{+10.3}_{-9.4}$ & $28.2^{+9.3}_{-10.1}$ & $16.0^{+7.4}_{-9.3}$ & $5.7^{+4.5}_{-4.7}$ \\
           \hdashline
            $50\%\ \rho_{\rm{UV}}\ (M_{\rm{UV}})$ & -17.3  & -17.0 & -16.7 &  -16.3 & -16.0  & -15.5  & -14.6  \\
            $90\%\ \rho_{\rm{UV}}\ (M_{\rm{UV}})$ & -12.8 &  -13.5 & -13.7 & -13.8 & -13.7 & -13.2 & -12.2 \\
            \hline
        \end{tabularx}
    \end{center}
    \caption{  The cosmic SFR density obtained by integrating our BDA LFs down to the commonly-chosen limit of $M_{\rm{UV}}<-17$ (equivalent to SFR $\gsim0.32\ \rm{M_{\odot}\ yr^{-1}}$) ({\it top row}), compared with the total (cumulative) SFRD ({\it second row}). The third row shows the resulting completeness. Errors correspond to 68\% C.L. 
      The bottom two rows denote the UV magnitude limit corresponding to 50\% and 90\% of the cumulative UV luminosity density (illustrated in Fig. \ref{Fig:BDA_frise} by the dashed black lines). }
\label{Tab:SFRD}
\end{table*}
\endgroup

We apply BDA on the four data sets in order to compare them and create a combined LF. As explained above, the relative evidence is computed from the 10 data points in the magnitude range $\rm{[-20,-15]}$ at redshift 6 for each data set. These data are illustrated in Fig. \ref{Fig:OBS_zoom}. Note that the ultra-faint end, where the difference between observational teams is maximal, is not used for the relative evidence. The restriction to this smaller range of magnitudes and redshifts does slightly impact the final posterior distribution, as discussed in Section \ref{Sec:Observations}, but it is necessary in order to apply the BDA method.

Table \ref{Tab:BDA_evidence} gives the resulting relative evidence of the data sets.  The I+ and A+ data sets are preferred by our model compared to the two others.  This preference is mostly due to the combination of (i) smoothness of the points over the reference range and (ii) small error
    \footnote{ As demonstrated in \S \ref{Sec:Demonstration on toy models}, errors which are too small are naturally penalized by BDA.  We can however explicitly check if the data sets have underestimated errors by computing their $\chi^2$ to the corresponding ML model.  The   resulting $\chi^2$ are 2.5 / 3.7 / 0.9 / 1.7 for B+ / I+ / A+ and L+.  Although A+ has the smallest chi-squared, it is consistent with a  $\chi^2$ distribution with three effective degrees of freedom (like our model). I+ has the largest chi-squared (within 71\% C.L. of the chi-squared distribution), which is even higher if one uses the quoted errors instead of the 20\% minimum errors that we applied ($\chi^2$ of 5.9 at 88\% C.L.).  This is weakly suggestive that the errors in the I+ data set could be underestimated.}
bars which are still consistent with our parametric model.  The L+ data-set is disfavored because it has a plateau at $M_{\rm UV}=$ -19.5 -- -18.5 and a steepening at the faint end; these features are difficult to fit with our model which relies on smooth functions on top of the HMFs. B+ also has small relative evidence, mainly because of the non-monotonic feature at  $M_{\rm UV}=$ -16, and the comparably large error bars at the bright end of the range.  

We can now combine the posteriors of each individual data sets, weighted by this relative evidence (eq. \ref{Eq:BDA_combining}).  We note that, although the relative weights are computed using only the ten LF points at $z=6$ common to every data set,  {\it each individual posterior is then re-computed using all the data available in the data set i.e. including the ultra-faint end and all redshifts (see Fig. \ref{Fig:all posterior}).}
It is these posteriors resulting from all data points which are averaged using the relative weights in table \ref{Tab:BDA_evidence}, resulting in the combined posteriors shown in Fig. \ref{Fig:posterior}.  To summarize, the weights are computed on comparable data, at redshift 6 in the magnitude range $\rm{[-20,-15]}$ and are applied on the posterior computed using all the data available.

There are several trends evident in Fig. \ref{Fig:posterior}.  Firstly, we note the degeneracy between $f_{\rm{*,10}}$ and $t_*$, as the ratio of the two  ($r_* = t_* / f_{\rm{*,10}}$) is relevant for the LFs (see Appendix \ref{App:The ratio}).  Following \citet{Park18}, we use a linear prior over log($f_{\rm{*,10}}$) and $\rm{t_*}$; as a result, the later is not constrained, showing a flat distribution over the full range.

The double peak in the 1D marginalized posterior of $\rm{\alpha_*}$ comes from the fact that the two data sets driving the combined posteriors (A+ and I+) favor two different values for this slope of the $M_\ast$ -- $M_{\rm h}$ relation (see Appendix \ref{App:Comparison of all posterior distributions}). A+ in particular favors a steeper LF (smaller $\alpha_\ast)$, resulting in a marginalized one sigma constraint of $\alpha_\ast = \rm{0.2^{+0.09}_{-0.07}}$. This can be understood since the data points that are most constraining are those with the smallest errors. For A+ as for I+, the error is minimum at the bright end of the range we use (see Fig. \ref{Fig:OBS_zoom}), and for A+ these points have a steeper slope.

The combined marginalized posterior also shows some constraints on $M_{\rm{t}}$, which peaks at $\rm{9.39^{+0.23}_{-1.35}\ [log_{10}(M_{\odot})]}$. This peak is entirely driven by A+ ($\rm{9.55^{+0.13}_{-0.55}\ [log_{10}(M_{\odot})]}$), with all of the other data sets only providing an upper limit (see Fig. D1).  However, the statistical significance of this peak is down-weighted by the BDA combined posteriors, resulting in only an upper limit on the turn-over scale (see also \citealt{Yue16}, where they look for evidence of a feedback-induced turn over in the LF).

\subsection{The combined luminosity functions}
\label{Sec:The combined Luminosity Function}

The posterior over the parameter space is sampled to obtain the corresponding constraints on the LFs.  In Fig. \ref{Fig:BDA_frise} we present the LF constraints corresponding to the 68\% C.L. range of the BDA posteriors {\it blue shaded areas}.  One nice result from this procedure is the forecast of LFs at even higher redshifts at which we currently have no data (c.f. $z=15$ LFs in the rightmost panel); although we caution that as our model is mostly constrained by the $z=6$ points, these extensions to higher redshifts are even more model-dependent.  We provide the numerical values for these LF constraints in tables \ref{Tab:LF678}, \ref{Tab:LF91012} and \ref{Tab:LF15}.

In this figure, we also compare the BDA LFs with those resulting from a uniform weighing of the estimated data sets, i.e. a simple average of each individual posterior, giving a relative weight of 25\% to all data sets.  The 68\% C.L. of the LFs obtained through this simple averaging are shown with the orange shaded regions in Fig. \ref{Fig:BDA_frise}.  Comparing the orange and the blue shaded regions, we see that the posteriors obtained with BDA are broader, allowing for a turn-over at brighter magnitudes.  This is driven by the fact that the A+ data set, which is the only one showing evidence of a turn-over, has a larger relative contribution in the BDA posterior (43\% compared to 25\%).  Specifically, we note that BDA LFs do not start to flatten or turn-over until {\it at least} $M_{\rm UV} \gsim -14$ (1 $\sigma$).  The corresponding scale is shifted fainter by 1 dex for the uniform weighted LFs, to $M_{\rm UV} \gsim -13$.

We can also compare our BDA combined LFs to those presented in \cite{Yue18}, who use redshift 6 blank field data from \cite{Bouwens15}, complemented with their own lensed galaxy estimates obtained by taking a mean probability of the number of galaxies per bin implied by different lensing models.  The resulting LFs are presented in terms of confidence limits, obtained by sampling a Schecter function modified to allow for a turn-over, and shown in the first panel of Fig. \ref{Fig:BDA_frise}.  
Their LFs at the bright end of the range are in agreement with our BDA combined LFs; however, their  68\% contours for magnitudes fainter than $M_{\rm UV} \gsim -15$ are broader than the ones resulting from BDA.

The corresponding {\it cumulative} UV luminosity densities for the BDA LFs are shown in the bottom row of Fig.  \ref{Fig:BDA_frise}, with the dotted lines denoting 50\% and 90\% of the total UV luminosity density (see also the bottom two rows on table \ref{Tab:SFRD}).  At redshift 6, galaxies brighter than -17.3 (-12.8) contribute to 50\% (90\%) of the total UV luminosity.
The 50\% limit magnitude increases with redshift, increasing the contribution of fainter galaxies in the total UV budget. But at the same time, the 90\% limit magnitude does not significantly evolve with redshift.

It is important to note that the distribution of the {\it ionizing} photon number density (relevant for reionization) is likely shifted even further towards fainter galaxies than the non-ionizing UV luminosity density.  This is because the ionizing escape fraction is expected to increase towards smaller, fainter galaxies, in which it is easier for feedback to evacuate low column density channels facilitating the escape of ionizing photons ( e.g. \citealt{Razoumov2010, Yajima2011, Ferrara2013, Paardekooper2015, Xu2016, Kimm2017} ).  Therefore, when it comes to the total ionizing photon budget, faint galaxies are likely even more important than implied by the 1500 \AA\ CDFs shown in the bottom row of Fig. \ref{Fig:BDA_frise}.

 \subsection{Star formation rate density}
 \label{Sec:Star formation rate density}
  
Finally, in Fig. \ref{Fig:SFRD} we show the cosmic star formation rate density (SFRD) from the BDA LFs presented in the previous figure.  The SFRD is shown for two integration limits, up to the magnitude of -17 (with 68\% C.L. in blue) and integrating over the whole population (68\% C.L. in orange).  We see that the SFRD up to a magnitude limit of -17 is consistent with observational estimates over the corresponding range (homogenized to correspond to the same limit according to \citealt{Oesch18}).  However, accounting for star formation in fainter galaxies implies a less rapid decrease going towards higher redshifts.  For example, the SFRD down to -17 drops by 3.5 dex going from redshifts 6--15, while the total SFRD only decreases by 2.5 dex over the same redshift interval.

We also quote the median and 68\% C.L. in table \ref{Tab:SFRD} for these two integration limits (two first rows) as well as the completeness expressed in percent of the total SFRD.  At redshift 6, galaxies brighter than magnitude -17 account for 60\% of the total SFRD.  However, this completeness drops rapidly as we go deeper into the EoR and cosmic dawn, becoming only 6\% at $z=15$.\footnote{The completeness is even lower at higher redshifts if there is a separate, transient population of molecularly-cooled galaxies.  We expect these molecularly-cooled galaxies to have different properties compared with the galaxies we observe at $z\lsim10$ (e.g. \citealt{Wise2014, So2014}), and the framework we use here does not allow for disparate galaxy populations.  We will return to this in future work, focused on the ultra-high redshifts in which such galaxies are expected to live. }

\section{Conclusions}
\label{Sec:Conclusions}

High redshift LFs provide an important constraint on galaxy formation in the first billion years of the Universe.  However, the observations are very challenging, with some estimates disagreeing significantly.

Here we present a simple framework, Bayesian Data-analysis Averaging (BDA), to combine different high-$z$ LF estimations.  The approach relies on a simple analytic model to encapsulate what we expect from LFs (i.e. smoothness and dependence on halo mass functions) while allowing flexibility to account for the unknown physics behind them.

In principle, there are two uses of BDA: (i) to infer "true" LFs from disparate data sets; and (ii) to allow disparate data sets to constrain a galaxy model.  For the former, the inferred "true" LFs depend on the parametrization of the galaxy model, which should hopefully be reasonable and flexible.  For the later, the galaxy model is already assumed: if one is not doing model selection, the parametrization is implicitly the "truth" and one is only interested in constraining the parameters with observations.  In this case, BDA provides a way of combining disparate LF estimates self-consistently using the same galaxy parameterization that one wishes to constrain.

We apply BDA on four data sets of high-$z$  (z $\geq$ 6), faint-end $M_{\rm UV}> -20$  LFs.  The resulting posteriors are mostly driven by two of the four data sets, showing a corresponding bimodality in the implied $M_\ast$ -- $M_{\rm halo}$ relation.  The combined posterior also shows very weak evidence of a turn-over at faint magnitudes, driven entirely by one data set.  

We provide the BDA LFs corresponding to our combined posteriors, which could be used to constrain similar galaxy formation models.  These LFs extend to high redshifts and faint objects, for which we currently have no data.  However, those extrapolations are model dependent, implicitly relying on our galaxy parameters being able to characterize the true LFs.  The approach we present can be applied to future data sets, such as those expected from $JWST$, as well as providing a framework for galaxy models to be informed by disparate data sets.

%%%%%%%%%%%%%%%%%%%%%%%%%%%%%%%%%%%%%%%%%%%%%%%%%%


\bibliographystyle{mnras}
\bibliography{bib.bib} % if your bibtex file is called example.bib

\begin{thebibliography}{}
\makeatletter
\relax
\def\mn@urlcharsother{\let\do\@makeother \do\$\do\&\do\#\do\^\do\_\do\%\do\~}
\def\mn@doi{\begingroup\mn@urlcharsother \@ifnextchar [ {\mn@doi@}
  {\mn@doi@[]}}
\def\mn@doi@[#1]#2{\def\@tempa{#1}\ifx\@tempa\@empty \href
  {http://dx.doi.org/#2} {doi:#2}\else \href {http://dx.doi.org/#2} {#1}\fi
  \endgroup}
\def\mn@eprint#1#2{\mn@eprint@#1:#2::\@nil}
\def\mn@eprint@arXiv#1{\href {http://arxiv.org/abs/#1} {{\tt arXiv:#1}}}
\def\mn@eprint@dblp#1{\href {http://dblp.uni-trier.de/rec/bibtex/#1.xml}
  {dblp:#1}}
\def\mn@eprint@#1:#2:#3:#4\@nil{\def\@tempa {#1}\def\@tempb {#2}\def\@tempc
  {#3}\ifx \@tempc \@empty \let \@tempc \@tempb \let \@tempb \@tempa \fi \ifx
  \@tempb \@empty \def\@tempb {arXiv}\fi \@ifundefined
  {mn@eprint@\@tempb}{\@tempb:\@tempc}{\expandafter \expandafter \csname
  mn@eprint@\@tempb\endcsname \expandafter{\@tempc}}}

\bibitem[\protect\citeauthoryear{{Atek}, {Richard}, {Kneib}  \&
  {Schaerer}}{{Atek} et~al.}{2018}]{Atek18}
{Atek} H.,  {Richard} J.,  {Kneib} J.-P.,   {Schaerer} D.,  2018, \mn@doi
  [\mnras] {10.1093/mnras/sty1820}, \href
  {http://cdsads.u-strasbg.fr/abs/2018MNRAS.479.5184A} {479, 5184}

\bibitem[\protect\citeauthoryear{{Barkana} \& {Loeb}}{{Barkana} \&
  {Loeb}}{2007}]{BarkanaLoeb2007}
{Barkana} R.,  {Loeb} A.,  2007, \mn@doi [Reports on Progress in Physics]
  {10.1088/0034-4885/70/4/R02}, \href
  {http://cdsads.u-strasbg.fr/abs/2007RPPh...70..627B} {70, 627}

\bibitem[\protect\citeauthoryear{{Behroozi} \& {Silk}}{{Behroozi} \&
  {Silk}}{2015}]{Behroozi2015}
{Behroozi} P.~S.,  {Silk} J.,  2015, \mn@doi [\apj]
  {10.1088/0004-637X/799/1/32}, \href
  {https://ui.adsabs.harvard.edu/abs/2015ApJ...799...32B} {799, 32}

\bibitem[\protect\citeauthoryear{{Behroozi}, {Wechsler}  \&
  {Conroy}}{{Behroozi} et~al.}{2013}]{Behroozi2013}
{Behroozi} P.~S.,  {Wechsler} R.~H.,   {Conroy} C.,  2013, \mn@doi [\apj]
  {10.1088/0004-637X/770/1/57}, \href
  {http://cdsads.u-strasbg.fr/abs/2013ApJ...770...57B} {770, 57}

\bibitem[\protect\citeauthoryear{{Bernal} \& {Peacock}}{{Bernal} \&
  {Peacock}}{2018}]{Bernal2018}
{Bernal} J.~L.,  {Peacock} J.~A.,  2018, \mn@doi [\jcap]
  {10.1088/1475-7516/2018/07/002}, \href
  {http://cdsads.u-strasbg.fr/abs/2018JCAP...07..002B} {7, 002}

\bibitem[\protect\citeauthoryear{{Bouwens} et~al.,}{{Bouwens}
  et~al.}{2012}]{Bouwens2012}
{Bouwens} R.~J.,  et~al., 2012, \mn@doi [\apj] {10.1088/0004-637X/754/2/83},
  \href {https://ui.adsabs.harvard.edu/abs/2012ApJ...754...83B} {754, 83}

\bibitem[\protect\citeauthoryear{{Bouwens} et~al.,}{{Bouwens}
  et~al.}{2014}]{Bouwens2014}
{Bouwens} R.~J.,  et~al., 2014, \mn@doi [\apj] {10.1088/0004-637X/793/2/115},
  \href {http://cdsads.u-strasbg.fr/abs/2014ApJ...793..115B} {793, 115}

\bibitem[\protect\citeauthoryear{{Bouwens} et~al.}{{Bouwens}
  et~al.}{2015}]{Bouwens15}
{Bouwens} R.~J.,  et~al., 2015, \mn@doi [\apj] {10.1088/0004-637X/803/1/34},
  \href {http://cdsads.u-strasbg.fr/abs/2015ApJ...803...34B} {803, 34}

\bibitem[\protect\citeauthoryear{{Bouwens} et~al.,}{{Bouwens}
  et~al.}{2016}]{Bouwens2016}
{Bouwens} R.~J.,  et~al., 2016, \mn@doi [\apj] {10.3847/1538-4357/833/1/72},
  \href {https://ui.adsabs.harvard.edu/abs/2016ApJ...833...72B} {833, 72}

\bibitem[\protect\citeauthoryear{{Bouwens}, {Oesch}, {Illingworth}, {Ellis}  \&
  {Stefanon}}{{Bouwens} et~al.}{2017}]{Bouwens17}
{Bouwens} R.~J.,  {Oesch} P.~A.,  {Illingworth} G.~D.,  {Ellis} R.~S.,
  {Stefanon} M.,  2017, \mn@doi [\apj] {10.3847/1538-4357/aa70a4}, \href
  {http://cdsads.u-strasbg.fr/abs/2017ApJ...843..129B} {843, 129}

\bibitem[\protect\citeauthoryear{{Bowman}, {Rogers}, {Monsalve}, {Mozdzen}  \&
  {Mahesh}}{{Bowman} et~al.}{2018}]{Bowman2018}
{Bowman} J.~D.,  {Rogers} A. E.~E.,  {Monsalve} R.~A.,  {Mozdzen} T.~J.,
  {Mahesh} N.,  2018, \mn@doi [\nat] {10.1038/nature25792}, \href
  {https://ui.adsabs.harvard.edu/abs/2018Natur.555...67B} {555, 67}

\bibitem[\protect\citeauthoryear{{Cullen}, {McLure}, {Khochfar}, {Dunlop}  \&
  {Dalla Vecchia}}{{Cullen} et~al.}{2017}]{Cullen2017}
{Cullen} F.,  {McLure} R.~J.,  {Khochfar} S.,  {Dunlop} J.~S.,   {Dalla
  Vecchia} C.,  2017, \mn@doi [\mnras] {10.1093/mnras/stx1451}, \href
  {https://ui.adsabs.harvard.edu/abs/2017MNRAS.470.3006C} {470, 3006}

\bibitem[\protect\citeauthoryear{{Dayal} \& {Ferrara}}{{Dayal} \&
  {Ferrara}}{2018}]{Dayal2018}
{Dayal} P.,  {Ferrara} A.,  2018, \mn@doi [\physrep]
  {10.1016/j.physrep.2018.10.002}, \href
  {http://cdsads.u-strasbg.fr/abs/2018PhR...780....1D} {780, 1}

\bibitem[\protect\citeauthoryear{{Dayal}, {Ferrara}, {Dunlop}  \&
  {Pacucci}}{{Dayal} et~al.}{2014}]{Dayal2014}
{Dayal} P.,  {Ferrara} A.,  {Dunlop} J.~S.,   {Pacucci} F.,  2014, \mn@doi
  [\mnras] {10.1093/mnras/stu1848}, \href
  {http://cdsads.u-strasbg.fr/abs/2014MNRAS.445.2545D} {445, 2545}

\bibitem[\protect\citeauthoryear{{Ferrara} \& {Loeb}}{{Ferrara} \&
  {Loeb}}{2013}]{Ferrara2013}
{Ferrara} A.,  {Loeb} A.,  2013, \mn@doi [\mnras] {10.1093/mnras/stt381}, \href
  {https://ui.adsabs.harvard.edu/abs/2013MNRAS.431.2826F} {431, 2826}

\bibitem[\protect\citeauthoryear{{Fialkov}, {Barkana}, {Visbal},
  {Tseliakhovich}  \& {Hirata}}{{Fialkov} et~al.}{2013}]{Fialkov2013}
{Fialkov} A.,  {Barkana} R.,  {Visbal} E.,  {Tseliakhovich} D.,   {Hirata}
  C.~M.,  2013, \mn@doi [\mnras] {10.1093/mnras/stt650}, \href
  {https://ui.adsabs.harvard.edu/abs/2013MNRAS.432.2909F} {432, 2909}

\bibitem[\protect\citeauthoryear{{Finkelstein}}{{Finkelstein}}{2016}]{Finkelstein2016}
{Finkelstein} S.~L.,  2016, \mn@doi [\pasa] {10.1017/pasa.2016.26}, \href
  {http://cdsads.u-strasbg.fr/abs/2016PASA...33...37F} {33, e037}

\bibitem[\protect\citeauthoryear{{Finkelstein} et~al.,}{{Finkelstein}
  et~al.}{2012}]{Finkelstein12}
{Finkelstein} S.~L.,  et~al., 2012, \mn@doi [\apj]
  {10.1088/0004-637X/758/2/93}, \href
  {https://ui.adsabs.harvard.edu/abs/2012ApJ...758...93F} {758, 93}

\bibitem[\protect\citeauthoryear{{Finlator} et~al.,}{{Finlator}
  et~al.}{2017}]{Finlator2017}
{Finlator} K.,  et~al., 2017, \mn@doi [\mnras] {10.1093/mnras/stw2433}, \href
  {https://ui.adsabs.harvard.edu/abs/2017MNRAS.464.1633F} {464, 1633}

\bibitem[\protect\citeauthoryear{{Hobson}, {Bridle}  \& {Lahav}}{{Hobson}
  et~al.}{2002}]{Hobson2002}
{Hobson} M.~P.,  {Bridle} S.~L.,   {Lahav} O.,  2002, \mn@doi [\mnras]
  {10.1046/j.1365-8711.2002.05614.x}, \href
  {http://cdsads.u-strasbg.fr/abs/2002MNRAS.335..377H} {335, 377}

\bibitem[\protect\citeauthoryear{{Holzbauer} \& {Furlanetto}}{{Holzbauer} \&
  {Furlanetto}}{2012}]{Holzbauer2012}
{Holzbauer} L.~N.,  {Furlanetto} S.~R.,  2012, \mn@doi [\mnras]
  {10.1111/j.1365-2966.2011.19752.x}, \href
  {https://ui.adsabs.harvard.edu/abs/2012MNRAS.419..718H} {419, 718}

\bibitem[\protect\citeauthoryear{{Ishigaki}, {Kawamata}, {Ouchi}, {Oguri},
  {Shimasaku}  \& {Ono}}{{Ishigaki} et~al.}{2018}]{Ishigaki18}
{Ishigaki} M.,  {Kawamata} R.,  {Ouchi} M.,  {Oguri} M.,  {Shimasaku} K.,
  {Ono} Y.,  2018, \mn@doi [\apj] {10.3847/1538-4357/aaa544}, \href
  {http://cdsads.u-strasbg.fr/abs/2018ApJ...854...73I} {854, 73}

\bibitem[\protect\citeauthoryear{{Kennicutt}}{{Kennicutt}}{1998}]{Kennicutt1998}
{Kennicutt} Robert~C. J.,  1998, \mn@doi [\araa]
  {10.1146/annurev.astro.36.1.189}, \href
  {https://ui.adsabs.harvard.edu/abs/1998ARA&A..36..189K} {36, 189}

\bibitem[\protect\citeauthoryear{{Kimm}, {Katz}, {Haehnelt}, {Rosdahl},
  {Devriendt}  \& {Slyz}}{{Kimm} et~al.}{2017}]{Kimm2017}
{Kimm} T.,  {Katz} H.,  {Haehnelt} M.,  {Rosdahl} J.,  {Devriendt} J.,   {Slyz}
  A.,  2017, \mn@doi [\mnras] {10.1093/mnras/stx052}, \href
  {https://ui.adsabs.harvard.edu/abs/2017MNRAS.466.4826K} {466, 4826}

\bibitem[\protect\citeauthoryear{{Koh} \& {Wise}}{{Koh} \&
  {Wise}}{2018}]{Koh2018}
{Koh} D.,  {Wise} J.~H.,  2018, \mn@doi [\mnras] {10.1093/mnras/stx3018}, \href
  {https://ui.adsabs.harvard.edu/abs/2018MNRAS.474.3817K} {474, 3817}

\bibitem[\protect\citeauthoryear{{Lahav}, {Bridle}, {Hobson}, {Lasenby}  \&
  {Sodr{\'e}}}{{Lahav} et~al.}{2000}]{Lahav2000}
{Lahav} O.,  {Bridle} S.~L.,  {Hobson} M.~P.,  {Lasenby} A.~N.,   {Sodr{\'e}}
  L.,  2000, \mn@doi [\mnras] {10.1046/j.1365-8711.2000.03633.x}, \href
  {http://adsabs.harvard.edu/abs/2000MNRAS.315L..45L} {315, L45}

\bibitem[\protect\citeauthoryear{{Liu}, {Mutch}, {Angel}, {Duffy}, {Geil},
  {Poole}, {Mesinger}  \& {Wyithe}}{{Liu} et~al.}{2016}]{Liu2016}
{Liu} C.,  {Mutch} S.~J.,  {Angel} P.~W.,  {Duffy} A.~R.,  {Geil} P.~M.,
  {Poole} G.~B.,  {Mesinger} A.,   {Wyithe} J.~S.~B.,  2016, \mn@doi [\mnras]
  {10.1093/mnras/stw1015}, \href
  {http://cdsads.u-strasbg.fr/abs/2016MNRAS.462..235L} {462, 235}

\bibitem[\protect\citeauthoryear{{Livermore}, {Finkelstein}  \&
  {Lotz}}{{Livermore} et~al.}{2017}]{Livermore17}
{Livermore} R.~C.,  {Finkelstein} S.~L.,   {Lotz} J.~M.,  2017, \mn@doi [\apj]
  {10.3847/1538-4357/835/2/113}, \href
  {http://cdsads.u-strasbg.fr/abs/2017ApJ...835..113L} {835, 113}

\bibitem[\protect\citeauthoryear{{Loeb} \& {Furlanetto}}{{Loeb} \&
  {Furlanetto}}{2013}]{LoebFurlanetto2013}
{Loeb} A.,  {Furlanetto} S.~R.,  2013, {The First Galaxies in the Universe}

\bibitem[\protect\citeauthoryear{{Ma} \& {Berndsen}}{{Ma} \&
  {Berndsen}}{2014}]{MA2014}
{Ma} Y.-Z.,  {Berndsen} A.,  2014, \mn@doi [Astronomy and Computing]
  {10.1016/j.ascom.2014.04.005}, \href
  {http://cdsads.u-strasbg.fr/abs/2014A%26C.....5...45M} {5, 45}

\bibitem[\protect\citeauthoryear{{Ma} et~al.,}{{Ma} et~al.}{2019}]{Ma2019}
{Ma} X.,  et~al., 2019, arXiv e-prints, \href
  {https://ui.adsabs.harvard.edu/abs/2019arXiv190210152M} {p. arXiv:1902.10152}

\bibitem[\protect\citeauthoryear{{Madau} \& {Dickinson}}{{Madau} \&
  {Dickinson}}{2014}]{Madau2014}
{Madau} P.,  {Dickinson} M.,  2014, \mn@doi [\araa]
  {10.1146/annurev-astro-081811-125615}, \href
  {https://ui.adsabs.harvard.edu/abs/2014ARA&A..52..415M} {52, 415}

\bibitem[\protect\citeauthoryear{{McLeod}, {McLure}  \& {Dunlop}}{{McLeod}
  et~al.}{2016}]{McLeod2016}
{McLeod} D.~J.,  {McLure} R.~J.,   {Dunlop} J.~S.,  2016, \mn@doi [\mnras]
  {10.1093/mnras/stw904}, \href
  {https://ui.adsabs.harvard.edu/abs/2016MNRAS.459.3812M} {459, 3812}

\bibitem[\protect\citeauthoryear{{Mesinger}}{{Mesinger}}{2016}]{MesingerBook2016}
{Mesinger} A.,  ed. 2016, {Understanding the Epoch of Cosmic Reionization}
  Astrophysics and Space Science Library Vol. 423,
  \mn@doi{10.1007/978-3-319-21957-8.
}

\bibitem[\protect\citeauthoryear{{Mirocha} \& {Furlanetto}}{{Mirocha} \&
  {Furlanetto}}{2019}]{Mirocha2019}
{Mirocha} J.,  {Furlanetto} S.~R.,  2019, \mn@doi [Monthly Notices of the Royal
  Astronomical Society] {10.1093/mnras/sty3260}, \href
  {https://ui.adsabs.harvard.edu/abs/2019MNRAS.483.1980M} {483, 1980}

\bibitem[\protect\citeauthoryear{{O'Shea}, {Wise}, {Xu}  \& {Norman}}{{O'Shea}
  et~al.}{2015a}]{OShea2015}
{O'Shea} B.~W.,  {Wise} J.~H.,  {Xu} H.,   {Norman} M.~L.,  2015a, \mn@doi
  [\apj] {10.1088/2041-8205/807/1/L12}, \href
  {https://ui.adsabs.harvard.edu/abs/2015ApJ...807L..12O} {807, L12}

\bibitem[\protect\citeauthoryear{{O'Shea}, {Wise}, {Xu}  \& {Norman}}{{O'Shea}
  et~al.}{2015b}]{O'Shea2015}
{O'Shea} B.~W.,  {Wise} J.~H.,  {Xu} H.,   {Norman} M.~L.,  2015b, \mn@doi
  [\apjl] {10.1088/2041-8205/807/1/L12}, \href
  {https://ui.adsabs.harvard.edu/abs/2015ApJ...807L..12O} {807, L12}

\bibitem[\protect\citeauthoryear{{Ocvirk} et~al.,}{{Ocvirk}
  et~al.}{2016}]{Ocvirk2016}
{Ocvirk} P.,  et~al., 2016, \mn@doi [\mnras] {10.1093/mnras/stw2036}, \href
  {https://ui.adsabs.harvard.edu/abs/2016MNRAS.463.1462O} {463, 1462}

\bibitem[\protect\citeauthoryear{{Ocvirk} et~al.,}{{Ocvirk}
  et~al.}{2018}]{Ocvirk2018}
{Ocvirk} P.,  et~al., 2018, arXiv e-prints, \href
  {https://ui.adsabs.harvard.edu/abs/2018arXiv181111192O} {p. arXiv:1811.11192}

\bibitem[\protect\citeauthoryear{{Oesch} et~al.,}{{Oesch}
  et~al.}{2013}]{Oesch2013}
{Oesch} P.~A.,  et~al., 2013, \mn@doi [\apj] {10.1088/0004-637X/773/1/75},
  \href {https://ui.adsabs.harvard.edu/abs/2013ApJ...773...75O} {773, 75}

\bibitem[\protect\citeauthoryear{{Oesch} et~al.,}{{Oesch}
  et~al.}{2014}]{Oesch2014}
{Oesch} P.~A.,  et~al., 2014, \mn@doi [\apj] {10.1088/0004-637X/786/2/108},
  \href {https://ui.adsabs.harvard.edu/abs/2014ApJ...786..108O} {786, 108}

\bibitem[\protect\citeauthoryear{{Oesch}, {Bouwens}, {Illingworth}, {Labb{\'e}}
   \& {Stefanon}}{{Oesch} et~al.}{2018}]{Oesch18}
{Oesch} P.~A.,  {Bouwens} R.~J.,  {Illingworth} G.~D.,  {Labb{\'e}} I.,
  {Stefanon} M.,  2018, \mn@doi [\apj] {10.3847/1538-4357/aab03f}, \href
  {http://cdsads.u-strasbg.fr/abs/2018ApJ...855..105O} {855, 105}

\bibitem[\protect\citeauthoryear{{Okamoto}, {Gao}  \& {Theuns}}{{Okamoto}
  et~al.}{2008}]{Okamoto2008}
{Okamoto} T.,  {Gao} L.,   {Theuns} T.,  2008, \mn@doi [\mnras]
  {10.1111/j.1365-2966.2008.13830.x}, \href
  {https://ui.adsabs.harvard.edu/abs/2008MNRAS.390..920O} {390, 920}

\bibitem[\protect\citeauthoryear{{Paardekooper}, {Khochfar}  \& {Dalla
  Vecchia}}{{Paardekooper} et~al.}{2015}]{Paardekooper2015}
{Paardekooper} J.-P.,  {Khochfar} S.,   {Dalla Vecchia} C.,  2015, \mn@doi
  [\mnras] {10.1093/mnras/stv1114}, \href
  {https://ui.adsabs.harvard.edu/abs/2015MNRAS.451.2544P} {451, 2544}

\bibitem[\protect\citeauthoryear{{Park}, {Mesinger}, {Greig}  \&
  {Gillet}}{{Park} et~al.}{2018}]{Park18}
{Park} J.,  {Mesinger} A.,  {Greig} B.,   {Gillet} N.,  2018, arXiv e-prints,
  \href {http://cdsads.u-strasbg.fr/abs/2018arXiv180908995P} {}

\bibitem[\protect\citeauthoryear{{Parkinson} \& {Liddle}}{{Parkinson} \&
  {Liddle}}{2013}]{Parkinson2013}
{Parkinson} D.,  {Liddle} A.~R.,  2013, \mn@doi [Statistical Analysis and Data
  Mining: The ASA Data Science Journal, Vol.~9, Issue 1, p.~3-14]
  {10.1002/sam.11179}, \href
  {http://cdsads.u-strasbg.fr/abs/2013SADM....6....3P} {6, 3}

\bibitem[\protect\citeauthoryear{{Planck Collaboration} et~al.,}{{Planck
  Collaboration} et~al.}{2018}]{Planckcosmo2018}
{Planck Collaboration} et~al., 2018, arXiv e-prints, \href
  {http://cdsads.u-strasbg.fr/abs/2018arXiv180706209P} {}

\bibitem[\protect\citeauthoryear{{Razoumov} \& {Sommer-Larsen}}{{Razoumov} \&
  {Sommer-Larsen}}{2010}]{Razoumov2010}
{Razoumov} A.~O.,  {Sommer-Larsen} J.,  2010, \mn@doi [\apj]
  {10.1088/0004-637X/710/2/1239}, \href
  {https://ui.adsabs.harvard.edu/abs/2010ApJ...710.1239R} {710, 1239}

\bibitem[\protect\citeauthoryear{{Schaerer}}{{Schaerer}}{2002}]{Schaerer2002}
{Schaerer} D.,  2002, \mn@doi [\aap] {10.1051/0004-6361:20011619}, \href
  {https://ui.adsabs.harvard.edu/abs/2002A&A...382...28S} {382, 28}

\bibitem[\protect\citeauthoryear{{Schechter}}{{Schechter}}{1976}]{Schechter1976}
{Schechter} P.,  1976, \mn@doi [Astrophysical Journal] {10.1086/154079}, \href
  {http://cdsads.u-strasbg.fr/abs/1976ApJ...203..297S} {203, 297}

\bibitem[\protect\citeauthoryear{{So}, {Norman}, {Reynolds}  \& {Wise}}{{So}
  et~al.}{2014}]{So2014}
{So} G.~C.,  {Norman} M.~L.,  {Reynolds} D.~R.,   {Wise} J.~H.,  2014, \mn@doi
  [\apj] {10.1088/0004-637X/789/2/149}, \href
  {https://ui.adsabs.harvard.edu/abs/2014ApJ...789..149S} {789, 149}

\bibitem[\protect\citeauthoryear{{Sobacchi} \& {Mesinger}}{{Sobacchi} \&
  {Mesinger}}{2013}]{Sobacchi2013}
{Sobacchi} E.,  {Mesinger} A.,  2013, \mn@doi [\mnras] {10.1093/mnras/stt693},
  \href {https://ui.adsabs.harvard.edu/abs/2013MNRAS.432.3340S} {432, 3340}

\bibitem[\protect\citeauthoryear{{Sobacchi} \& {Mesinger}}{{Sobacchi} \&
  {Mesinger}}{2014}]{Sobacchi2014}
{Sobacchi} E.,  {Mesinger} A.,  2014, \memsai, \href
  {https://ui.adsabs.harvard.edu/abs/2014MmSAI..85..509S} {85, 509}

\bibitem[\protect\citeauthoryear{{Steidel}, {Adelberger}, {Giavalisco},
  {Dickinson}  \& {Pettini}}{{Steidel} et~al.}{1999}]{Steidel1999}
{Steidel} C.~C.,  {Adelberger} K.~L.,  {Giavalisco} M.,  {Dickinson} M.,
  {Pettini} M.,  1999, \mn@doi [\apj] {10.1086/307363}, \href
  {http://cdsads.u-strasbg.fr/abs/1999ApJ...519....1S} {519, 1}

\bibitem[\protect\citeauthoryear{{Sun} \& {Furlanetto}}{{Sun} \&
  {Furlanetto}}{2016}]{Sun17}
{Sun} G.,  {Furlanetto} S.~R.,  2016, \mn@doi [\mnras] {10.1093/mnras/stw980},
  \href {http://cdsads.u-strasbg.fr/abs/2016MNRAS.460..417S} {460, 417}

\bibitem[\protect\citeauthoryear{{Trotta}}{{Trotta}}{2008}]{Trotta2008}
{Trotta} R.,  2008, \mn@doi [Contemporary Physics] {10.1080/00107510802066753},
  \href {http://cdsads.u-strasbg.fr/abs/2008ConPh..49...71T} {49, 71}

\bibitem[\protect\citeauthoryear{{Tumlinson} \& {Shull}}{{Tumlinson} \&
  {Shull}}{2000}]{Tumlinson2000}
{Tumlinson} J.,  {Shull} J.~M.,  2000, \mn@doi [\apjl] {10.1086/312432}, \href
  {https://ui.adsabs.harvard.edu/abs/2000ApJ...528L..65T} {528, L65}

\bibitem[\protect\citeauthoryear{Wallis}{Wallis}{2014}]{wallis2014}
Wallis K.~F.,  2014, \mn@doi [Statist. Sci.] {10.1214/13-STS417}, 29, 106

\bibitem[\protect\citeauthoryear{{Wilkins}, {Bouwens}, {Oesch}, {Labb{\'e}},
  {Sargent}, {Caruana}, {Wardlow}  \& {Clay}}{{Wilkins}
  et~al.}{2016}]{Wilkins2016}
{Wilkins} S.~M.,  {Bouwens} R.~J.,  {Oesch} P.~A.,  {Labb{\'e}} I.,  {Sargent}
  M.,  {Caruana} J.,  {Wardlow} J.,   {Clay} S.,  2016, \mn@doi [\mnras]
  {10.1093/mnras/stv2263}, \href
  {https://ui.adsabs.harvard.edu/abs/2016MNRAS.455..659W} {455, 659}

\bibitem[\protect\citeauthoryear{{Wilkins}, {Feng}, {Di Matteo}, {Croft},
  {Lovell}  \& {Waters}}{{Wilkins} et~al.}{2017}]{Wilkins2017}
{Wilkins} S.~M.,  {Feng} Y.,  {Di Matteo} T.,  {Croft} R.,  {Lovell} C.~C.,
  {Waters} D.,  2017, \mn@doi [\mnras] {10.1093/mnras/stx841}, \href
  {https://ui.adsabs.harvard.edu/abs/2017MNRAS.469.2517W} {469, 2517}

\bibitem[\protect\citeauthoryear{{Wise}, {Demchenko}, {Halicek}, {Norman},
  {Turk}, {Abel}  \& {Smith}}{{Wise} et~al.}{2014}]{Wise2014}
{Wise} J.~H.,  {Demchenko} V.~G.,  {Halicek} M.~T.,  {Norman} M.~L.,  {Turk}
  M.~J.,  {Abel} T.,   {Smith} B.~D.,  2014, \mn@doi [\mnras]
  {10.1093/mnras/stu979}, \href
  {https://ui.adsabs.harvard.edu/abs/2014MNRAS.442.2560W} {442, 2560}

\bibitem[\protect\citeauthoryear{{Xu}, {Norman}, {O'Shea}  \& {Wise}}{{Xu}
  et~al.}{2016a}]{Xu2016}
{Xu} H.,  {Norman} M.~L.,  {O'Shea} B.~W.,   {Wise} J.~H.,  2016a, \mn@doi
  [\apj] {10.3847/0004-637X/823/2/140}, \href
  {https://ui.adsabs.harvard.edu/abs/2016ApJ...823..140X} {823, 140}

\bibitem[\protect\citeauthoryear{{Xu}, {Wise}, {Norman}, {Ahn}  \&
  {O'Shea}}{{Xu} et~al.}{2016b}]{Xu2016b}
{Xu} H.,  {Wise} J.~H.,  {Norman} M.~L.,  {Ahn} K.,   {O'Shea} B.~W.,  2016b,
  \mn@doi [\apj] {10.3847/1538-4357/833/1/84}, \href
  {https://ui.adsabs.harvard.edu/abs/2016ApJ...833...84X} {833, 84}

\bibitem[\protect\citeauthoryear{{Yajima}, {Choi}  \& {Nagamine}}{{Yajima}
  et~al.}{2011}]{Yajima2011}
{Yajima} H.,  {Choi} J.-H.,   {Nagamine} K.,  2011, \mn@doi [\mnras]
  {10.1111/j.1365-2966.2010.17920.x}, \href
  {https://ui.adsabs.harvard.edu/abs/2011MNRAS.412..411Y} {412, 411}

\bibitem[\protect\citeauthoryear{{Yoshida}, {Omukai}  \& {Hernquist}}{{Yoshida}
  et~al.}{2008}]{Yoshida2008}
{Yoshida} N.,  {Omukai} K.,   {Hernquist} L.,  2008, \mn@doi [Science]
  {10.1126/science.1160259}, \href
  {https://ui.adsabs.harvard.edu/abs/2008Sci...321..669Y} {321, 669}

\bibitem[\protect\citeauthoryear{{Yue}, {Ferrara}  \& {Xu}}{{Yue}
  et~al.}{2016}]{Yue16}
{Yue} B.,  {Ferrara} A.,   {Xu} Y.,  2016, \mn@doi [\mnras]
  {10.1093/mnras/stw2145}, \href
  {http://adsabs.harvard.edu/abs/2016MNRAS.463.1968Y} {463, 1968}

\bibitem[\protect\citeauthoryear{{Yue} et~al.,}{{Yue} et~al.}{2018}]{Yue18}
{Yue} B.,  et~al., 2018, \mn@doi [\apj] {10.3847/1538-4357/aae77f}, \href
  {https://ui.adsabs.harvard.edu/abs/2018ApJ...868..115Y} {868, 115}

\bibitem[\protect\citeauthoryear{{Yung}, {Somerville}, {Finkelstein}, {Popping}
   \& {Dav{\'e}}}{{Yung} et~al.}{2019}]{Yung19}
{Yung} L.~Y.~A.,  {Somerville} R.~S.,  {Finkelstein} S.~L.,  {Popping} G.,
  {Dav{\'e}} R.,  2019, \mn@doi [\mnras] {10.1093/mnras/sty3241}, \href
  {https://ui.adsabs.harvard.edu/abs/2019MNRAS.483.2983Y} {483, 2983}

\makeatother
\end{thebibliography}

%%%%%%%%%%%%%%%%%%%%%%%%%%%%%%%%%%%%%%%%%%%%%%%%%%

%%%%%%%%%%%%%%% ACKNOWLEDGEMENTS %%%%%%%%%%%%%%%%%%%

\section*{Acknowledgements}
We thank R. Bouwens, S. Finkelstein, and H. Atek for valuable comments on a draft version of this work.  We also thank R. Livermore for providing us with the unpublished, Eddington bias-corrected data points and P. Oesch for sharing with us the homogenized star formation rate density data points. 
This work was supported by the European Research Council (ERC) under the European Union's Horizon 2020 research and innovation programme (grant agreement No 638809 -- AIDA -- PI: Mesinger). The results presented here reflect the authors' views; the ERC is not responsible for their use.
We thank contributors to SciPy\footnote{http://www.scipy.org}, Matplotlib\footnote{http://www.matplotlib.sourceforge.net}, pyDOE\footnote{https://pythonhosted.org/pyDOE/}, and the Python programming language\footnote{http://www.python.org}.

%%%%%%%%%%%%%%%%%%%%%%%%%%%%%%%%%%%%%%%%%%%%%%%%%%

%%%%%%%%%%%%%%%%% APPENDICES %%%%%%%%%%%%%%%%%%%%%

\appendix

%%%%%%%%%%%%%%%%%%%%%%
\section{Split Norm}
\label{App:Split Norm}
%%%%%%%%%%%%%%%%%%%%%%

To take into account the asymmetric errors provided in the observations we used the split norm distribution \citep{wallis2014}.  It is just the concatenation of two half-normal distributions, re-normalized to ensure continuity at the origin:

\begin{equation}
    \rm{ \mathcal{S}(x) = \left\{
    \begin{array}{ll}
        \rm{ A exp\left( -\frac{1}{2} \frac{(x-\mu)^2}{\sigma_1^2} \right), x\leq \mu, }\\
        
        \rm{ A exp\left( -\frac{1}{2} \frac{(x-\mu)^2}{\sigma_2^2} \right), x\geq \mu, } \\
    
        A = ( {\sqrt{2\pi} \left( \frac{\sigma_1+\sigma_2}{2} \right)^2 } )^{-1}
    \end{array}
\right. }
\end{equation}

\begin{figure}
\includegraphics[width=\columnwidth]{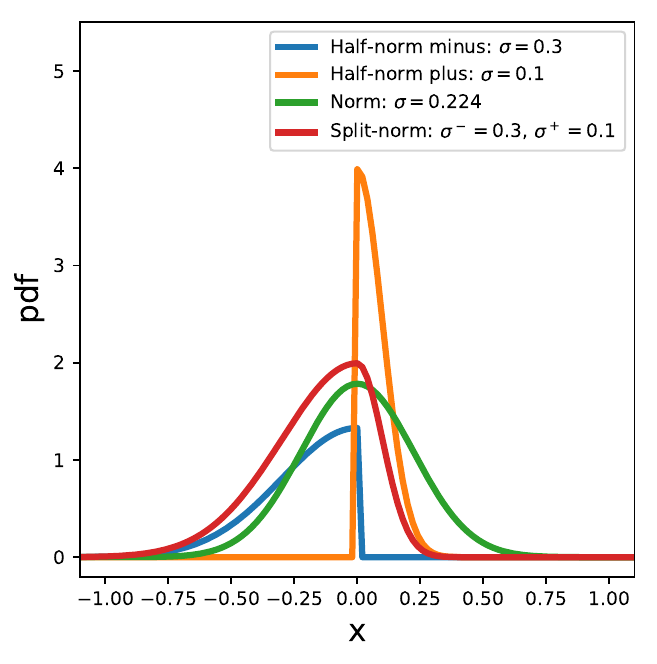}
\caption{ Example of application of the split normal distribution.  Two half normal distributions are shown in blue and orange with two different standard deviations (respectively 0.30 and 0.10).
The corresponding split-norm distribution is shown in red.  For comparison, we also show in green the normal distribution obtained using the average of the variance of the two half normal distributions (i.e. a standard deviation of $\sim$ 0.224).}
\label{Fig:splitNorm}
\end{figure}

For illustrative purposes, Fig. \ref{Fig:splitNorm} presents two half normal distributions in blue and orange with two different standard deviations (respectively 0.30 and 0.10).
The corresponding split-norm distribution is shown in red.  For comparison, we also show in green the normal distribution obtained using the average of the variance of the two half normal distributions (i.e. a standard deviation of $\sim$ 0.224).

%%%%%%%%%%%%%%%%%%%%%%%%%%%%%%%%%%%%%%%%%%%%%%%%%%%%%%%%%%%%%%
\section{Conversion of logarithmic to linear scale for errors}
\label{App:Conversion logarithm to linear scale of the errors}
%%%%%%%%%%%%%%%%%%%%%%%%%%%%%%%%%%%%%%%%%%%%%%%%%%%%%%%%%%%%%%

Some studies give the estimated data points and errors in logarithmic base 10 while others do so in linear scale.  In this study, we chose to work in logarithmic base 10. The transformation from linear to logarithmic scale for the errors are made as follows: 
\begin{equation}
    \rm{ \left\{
    \begin{array}{ll}
        \rm{ \phi_{log} = log_{10}(\phi_{lin}), }\\
        \rm{ \sigma_{log}^+ = log_{10}(\phi_{lin}+\sigma_{lin}^+) - log_{10}(\phi_{lin}), } \\
        \rm{ \sigma_{log}^- = log_{10}(\phi_{lin}) - log_{10}(\phi_{lin}-\sigma_{lin}^-), }
    \end{array}
    \right. }
\end{equation}
\begin{equation}
    \rm{ \left\{
    \begin{array}{ll}
        \rm{ \phi_{lin} = 10^{\phi_{log}}, }\\
        \rm{ \sigma_{lin}^+ = 10^{\phi_{log}+\sigma_{log}^+} - 10^{\phi_{log}}, } \\
        \rm{ \sigma_{lin}^- = 10^{\phi_{log}} - 10^{\phi_{log}-\sigma_{log}^-}. }
    \end{array}
    \right. }
\end{equation}

Note that symmetric errors in one scale become asymmetric in the other.  

%%%%%%%%%%%%%%%%%%%%%%%%%%%%
\section{The ratio $\lowercase{t_* / f_{\rm{*,10}}}$}
\label{App:The ratio}
%%%%%%%%%%%%%%%%%%%%%%%%%%%%

\begin{figure*}
\includegraphics[width=0.8\textwidth]{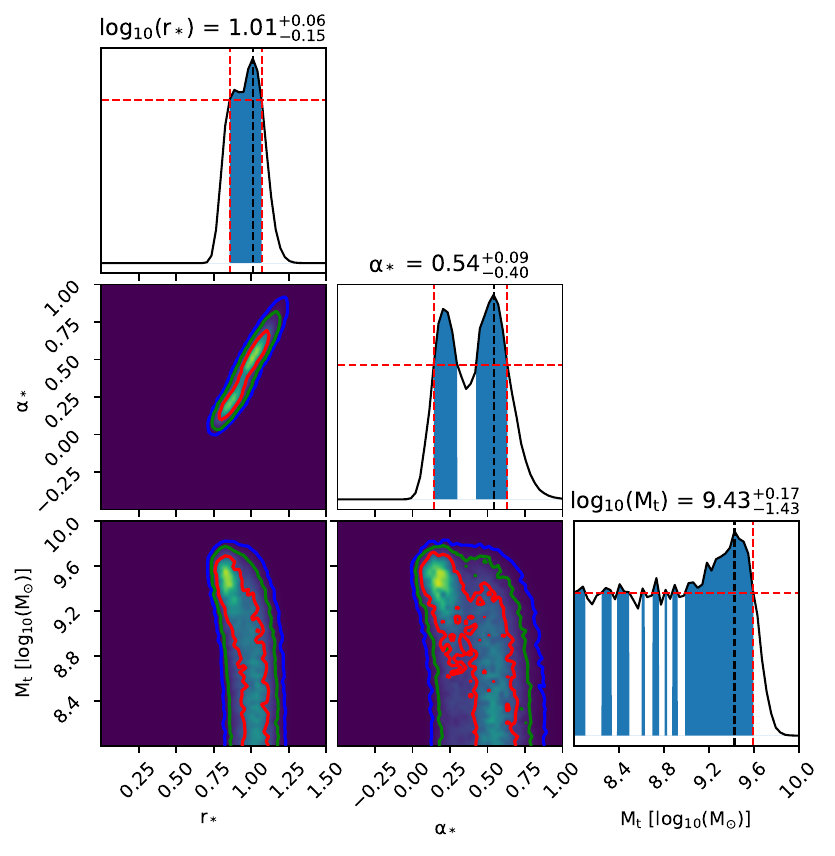}
\caption{ 1D and 2D marginalized posterior distributions of galaxy parameters resulting from the BDA weighing of the posteriors from each data set. The relative weights are listed in table \ref{Tab:BDA_evidence}. This figure is the same as Fig. \ref{Fig:posterior}, but here the degenerate parameters $f_{\rm{*,10}}$ and $t_*$ are replaced by their ratio $r_* = t_* / f_{\rm{*,10}}$ (in log scale). Note that the range of the ratio is zoomed, the original one derive from $f_{\rm{*,10}}$ and $t_*$ should be $[-4, 2.5]$. }
\label{Fig:posterior_ratio}
\end{figure*}

 The model used in this study contains two parameters that are completely degenerate in predicting the LF. Although only the ratio
  $r_* = t_* / f_{\rm{*,10}}$ is relevant for the LF, we explore the more general formulation by default in this work since EoR observations (or other data sets) can break this degeneracy (see \citealt{Park18}). 

In Fig. \ref{Fig:posterior_ratio} we replace $f_{\rm{*,10}}$ and $t_*$ by $r_*$ in the traditional corner plot of the posterior. It is the same posterior as presented in Fig. \ref{Fig:posterior}, i.e. it is derived from the BDA combination of the LF estimations. This ratio is strongly constrained by the LFs estimation, $\rm{log_{10}(r_*)=1.01_{-0.15}^{+0.06}  }$. It is degenerate with $\alpha_*$ and also slightly with $M_{\rm{t}}$ at large values of the latter. 
It is also noticeable that the sampling noise is reduced, due to the reduction of the parameter space dimensionality.

%%%%%%%%%%%%%%%%%%%%%%%%%%%%
\section{Convergence test}
\label{App:Convergence test}
%%%%%%%%%%%%%%%%%%%%%%%%%%%%

\begin{figure*}
\includegraphics[width=\textwidth]{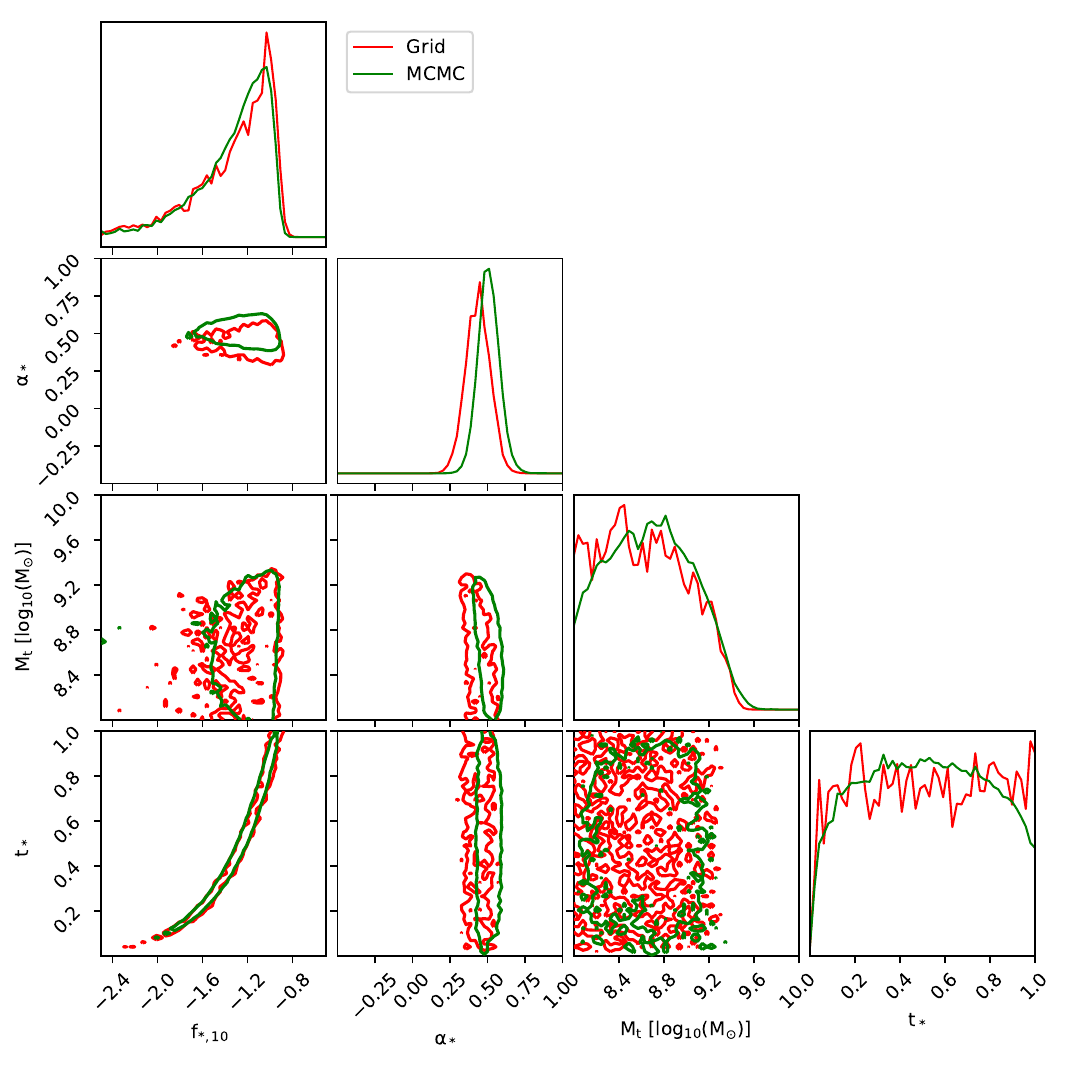}
\caption{ Comparison of the posterior distribution obtained with an on-the-fly MCMC and the pre-computed grid sampling of this study.  Both the MCMC and the grid contain the same number of samples: 200.000 points.  Those posteriors are generated using the B+ data set. The slight shift in the marginalized 1D constraints on $\alpha_\ast$ is due to the different treatment of errors, as discussed in the text.}
\label{Fig:convergence}
\end{figure*}

In this study, the likelihood is estimated on a grid of points sampled by LHS (200000 points).  To test the convergence of our estimation of the posterior distribution we compare it with the posterior distribution generated with on-the-fly MCMC sampling.  Note that the MCMC chain also contains 200000 points and has converged.  Fig. \ref{Fig:convergence} presents the comparison of the marginalized posterior distributions obtained with the grid (red) and with MCMC (green).  Both posteriors are generated using the B+ data set.  The 2D contour is the marginalized one sigma.  As expected, the marginalized distributions obtained using the grid sampling are noisier, but the final constraints are comparable.
We note a slight shift on the estimation of the parameter $\rm{ \alpha_* }$, due mostly to the difference in the treatment of the error:  for computational simplicity, the MCMC code used \citep{Park18} treats LF error bars as symmetric, while here we allow for asymmetry (see A1).

We perform a second convergence test on the estimation of the evidences, by comparing the results obtain with a sample 200000 points and one of 400000 points. The results are identical in both case proving the convergence of the estimation of the evidence.

%%%%%%%%%%%%%%%%%%%%%%%%%%%%%%%%%%%%%%
\section{Comparison of all posterior distributions}
\label{App:Comparison of all posterior distributions}
%%%%%%%%%%%%%%%%%%%%%%%%%%%%%%%%%%%%%%

\begin{figure*}
\includegraphics[width=\textwidth]{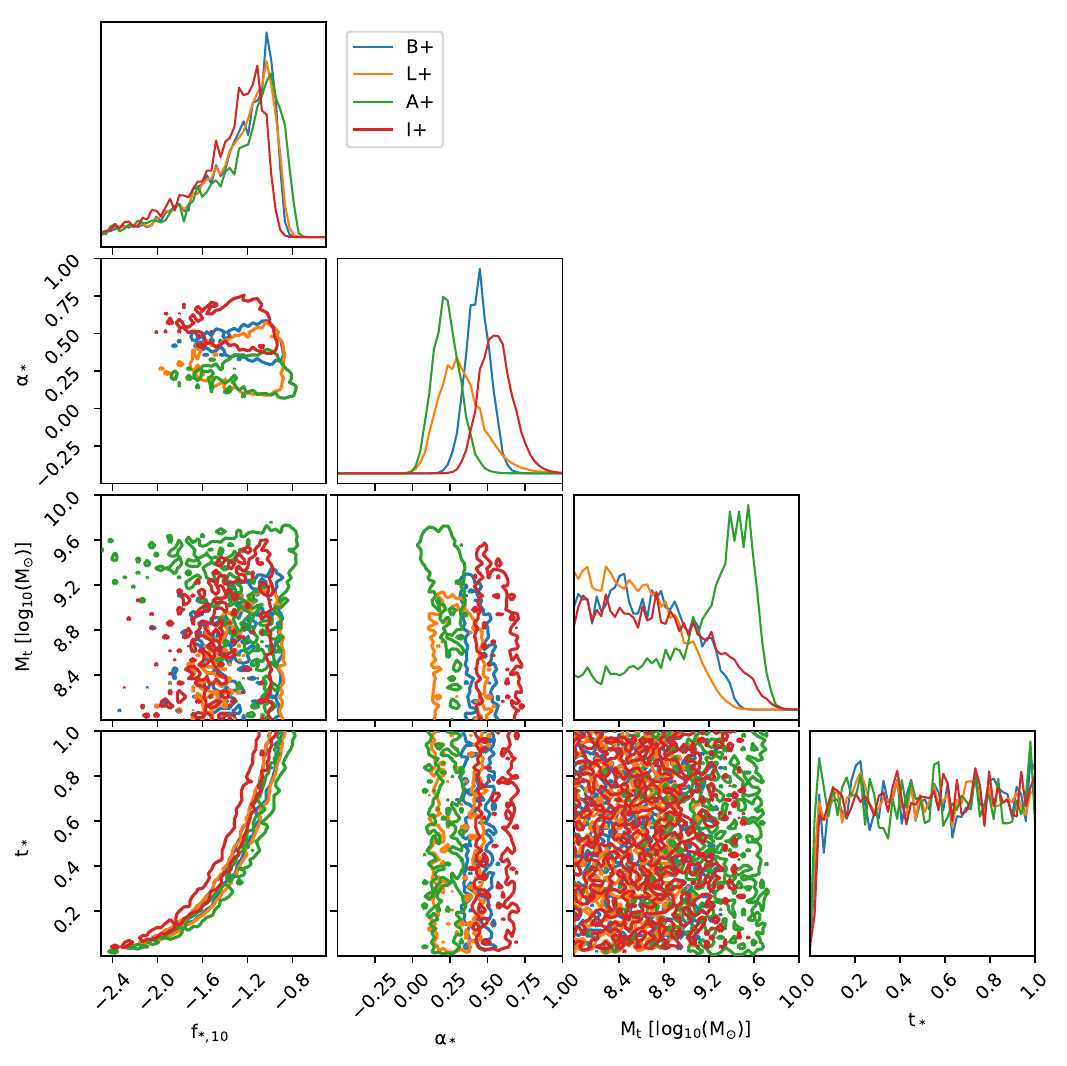}
\caption{ Comparison of the posterior distribution obtained with the four estimated data sets, using all the data available at every redshift. }
\label{Fig:all posterior}
\end{figure*}

We compare the posterior distribution obtained with the four data set in Fig. \ref{Fig:all posterior}.  These are generated using all data points for data sets. While Fig. present the same comparison but only using the 10 data points at redshift 6 with -20 $\leq M_{\rm{UV}} \leq$ -15.  

\begin{figure*}
\includegraphics[width=\textwidth]{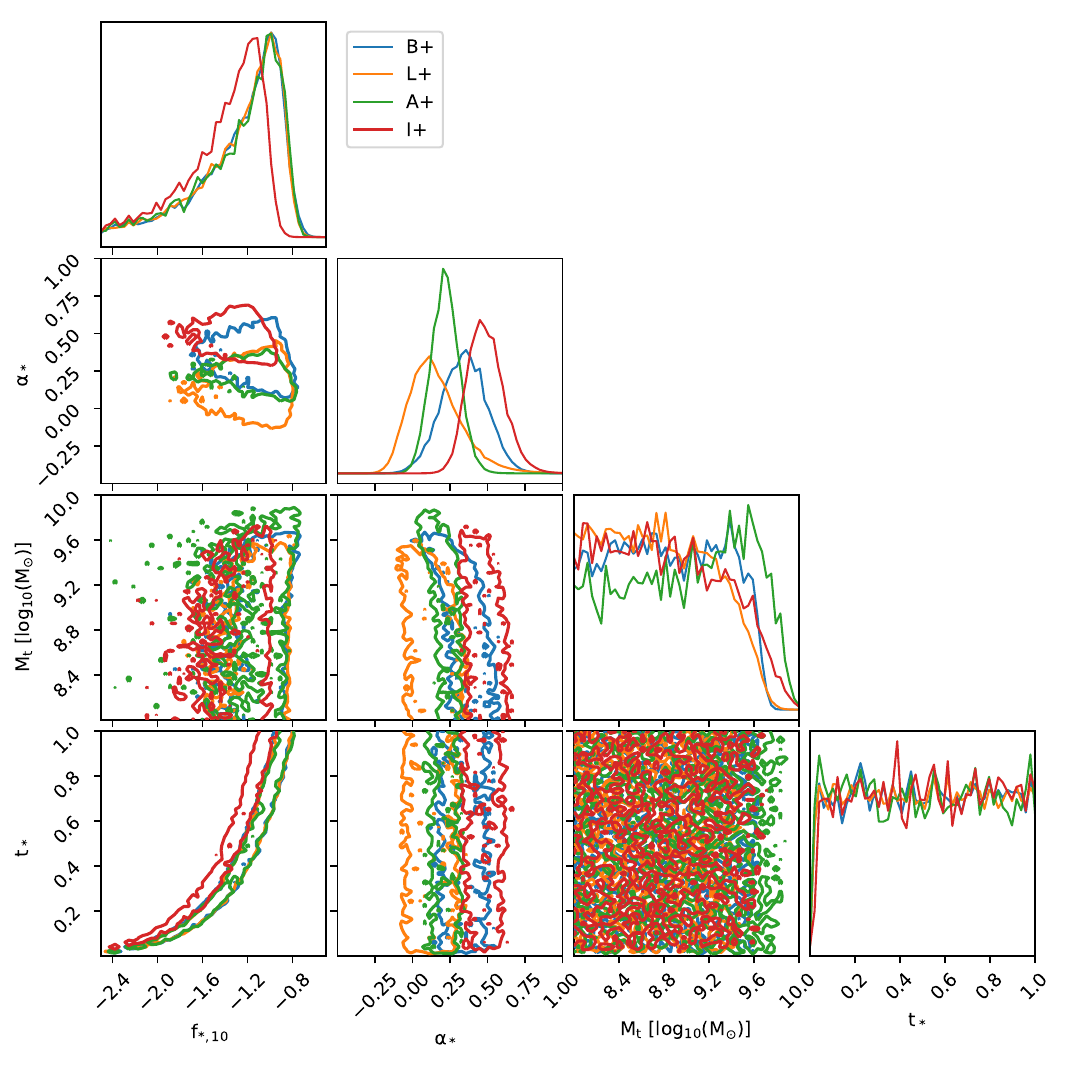}
\caption{ Comparison of the posterior distribution obtained with the four estimated data sets, using only the 10 data points at redshift 6 with -20 $\leq M_{\rm{UV}} \leq$ -15. }
\label{Fig:all posterior reduced data }
\end{figure*}

%%%%%%%%%%%%%%%%%%%%%%%%%%%%%%%%%%%%%%
\section{Comparison BDA, Average and reduce data sets}
\label{App:Comparison BDA and Average}
%%%%%%%%%%%%%%%%%%%%%%%%%%%%%%%%%%%%%%

\begin{figure*}
\includegraphics[width=\textwidth]{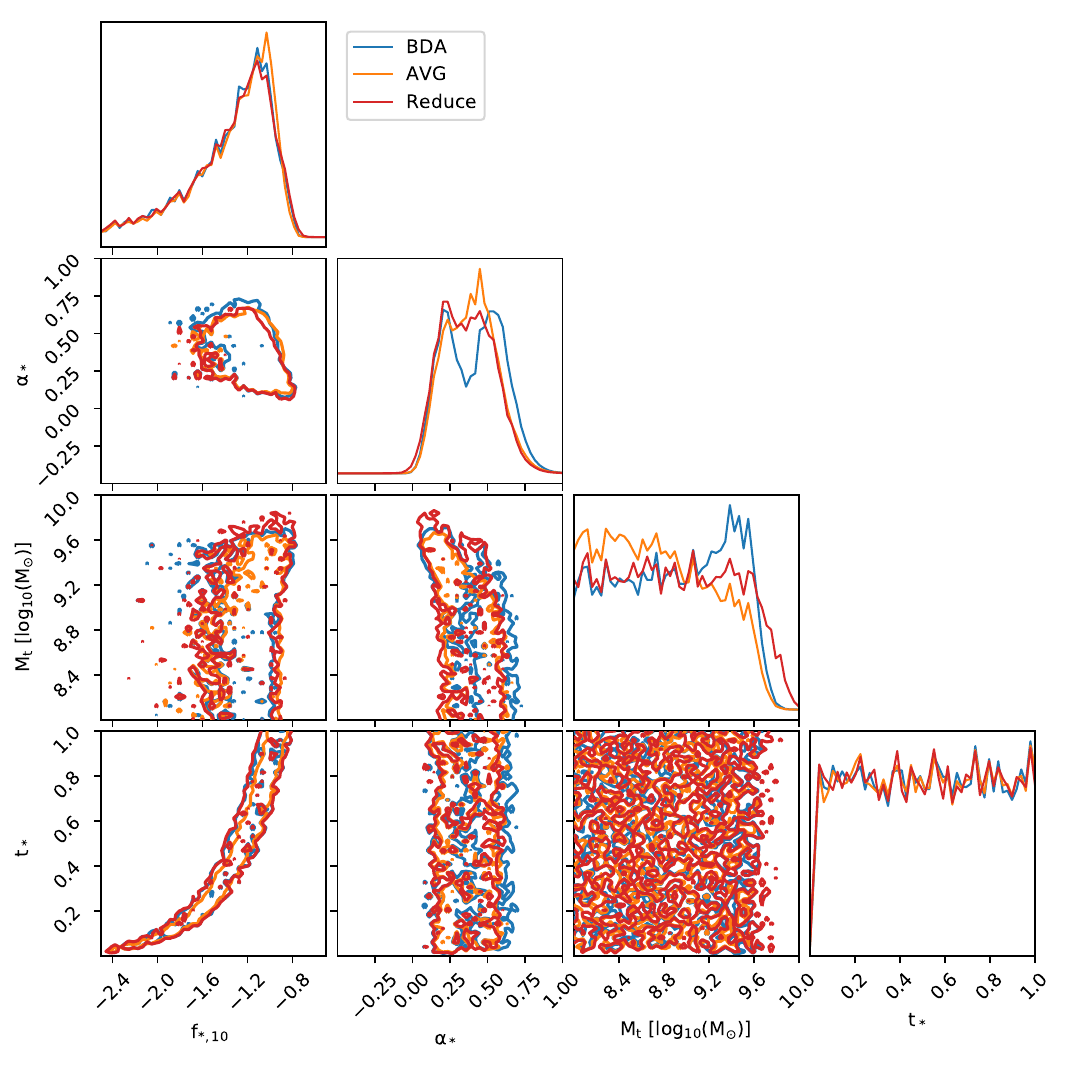}
\caption{ Comparison of the combined posterior distribution obtained with our fiducial BDA procedure ({\it blue}), a simple average of the individual posteriors ({\it orange}), and the BDA posterior obtained using only the data range used for computing the evidence ({\it red}; i.e. using only the redshift 6 points in the magnitude range [-20,-15].  The reduced range and the simple average both decrease the evidence for a turnover and the related bimodality in $\alpha_\ast$, as these trends are driven by the ultra faint, $M_{\rm uv}>-15$, points from the A+ data set.}
\label{Fig:Comparison BDA AVG}
\end{figure*}

We compare the posterior distribution obtained with the BDA method with a simple average of all individual posterior.  Fig. \ref{Fig:Comparison BDA AVG} present in blue the BDA posterior and in orange the average posterior.  There are two noticeable differences.  The first is on the parameter $\alpha_*$, in the average case, the distribution has a more Gaussian shape.  But this difference has no noticeable effect once projected on the LF space (see Fig \ref{Fig:BDA_frise}).  The second difference is the parameter $M_{\rm{t}}$:
in the case of the average, it is just a lower limit.  This effect is visible in the LF space (c.f. Fig. \ref{Fig:frise_OBS} and associated discussion).

The posterior distribution using only the redshift 6 data in the magnitude range [-20,-15] is presented by the red curve.  The bimodality of $\alpha_*$ is smoothed, because the constraints of this parameters are wider with the reduce data and the estimation of I+ is slightly smaller.  The second difference is for $M_{\rm{t}}$, where the peak at 9.4 is reduce, because it is driven by the ultra-faint end.

%%%%%%%%%%%%%%%%%%%%%%%%%%%%%%%%%%%%%%
\section{Tables of Luminosity Functions}
\label{App:Tables of Luminosity Functions}
%%%%%%%%%%%%%%%%%%%%%%%%%%%%%%%%%%%%%%

\renewcommand{\tabularxcolumn}[1]{>{\centering\arraybackslash}m{#1}}
\begin{table*}
    \begin{center}
        \begin{tabularx}{\textwidth}{XXXXXXXXXX}
            \hline
              & \multicolumn{3}{|c}{$z=6$} & \multicolumn{3}{|c}{$z=7$}  & \multicolumn{3}{|c|}{$z=8$} \\
              \hline
            $M_{\rm{UV}}$ & \multicolumn{1}{|c}{$\phi$} & $\sigma_{sup}$  & $\sigma_{inf}$ & \multicolumn{1}{|c}{$\phi$} & $\sigma_{sup}$  & $\sigma_{inf}$  & \multicolumn{1}{|c}{$\phi$} & $\sigma_{sup}$ & \multicolumn{1}{c|}{$\sigma_{inf}$}  \\
            \hline
-20.11 & -3.26 & 0.15 & 0.12 & -3.57 & 0.15 & 0.13 & -3.95 & 0.17 & 0.14 \\
-19.77 & -3.09 & 0.12 & 0.10 & -3.38 & 0.12 & 0.10 & -3.73 & 0.13 & 0.11 \\
-19.43 & -2.92 & 0.09 & 0.08 & -3.20 & 0.10 & 0.08 & -3.53 & 0.10 & 0.09 \\
-19.09 & -2.77 & 0.08 & 0.07 & -3.02 & 0.08 & 0.07 & -3.33 & 0.08 & 0.08 \\
-18.75 & -2.61 & 0.07 & 0.06 & -2.86 & 0.07 & 0.07 & -3.14 & 0.08 & 0.08 \\
-18.41 & -2.47 & 0.06 & 0.06 & -2.69 & 0.07 & 0.07 & -2.96 & 0.07 & 0.09 \\
-18.07 & -2.32 & 0.07 & 0.07 & -2.53 & 0.07 & 0.09 & -2.79 & 0.09 & 0.09 \\
-17.73 & -2.19 & 0.07 & 0.08 & -2.38 & 0.08 & 0.10 & -2.62 & 0.09 & 0.11 \\
-17.39 & -2.05 & 0.09 & 0.09 & -2.23 & 0.10 & 0.11 & -2.45 & 0.11 & 0.12 \\
-17.05 & -1.92 & 0.10 & 0.11 & -2.09 & 0.11 & 0.12 & -2.30 & 0.13 & 0.14 \\
-16.71 & -1.79 & 0.11 & 0.12 & -1.95 & 0.12 & 0.13 & -2.15 & 0.14 & 0.14 \\
-16.37 & -1.67 & 0.12 & 0.12 & -1.82 & 0.13 & 0.14 & -2.01 & 0.15 & 0.15 \\
-16.03 & -1.56 & 0.12 & 0.13 & -1.70 & 0.14 & 0.14 & -1.88 & 0.15 & 0.16 \\
-15.69 & -1.45 & 0.13 & 0.14 & -1.59 & 0.14 & 0.15 & -1.76 & 0.16 & 0.16 \\
-15.35 & -1.36 & 0.14 & 0.14 & -1.49 & 0.14 & 0.16 & -1.66 & 0.16 & 0.17 \\
-15.01 & -1.27 & 0.14 & 0.15 & -1.40 & 0.15 & 0.17 & -1.56 & 0.17 & 0.19 \\
-14.67 & -1.19 & 0.15 & 0.17 & -1.32 & 0.17 & 0.18 & -1.47 & 0.20 & 0.19 \\
-14.33 & -1.12 & 0.19 & 0.18 & -1.25 & 0.22 & 0.19 & -1.40 & 0.26 & 0.21 \\
-13.99 & -1.06 & 0.24 & 0.20 & -1.17 & 0.28 & 0.22 & -1.32 & 0.32 & 0.25 \\
-13.65 & -0.99 & 0.31 & 0.23 & -1.11 & 0.35 & 0.28 & -1.25 & 0.41 & 0.30 \\
-13.31 & -0.94 & 0.40 & 0.27 & -1.05 & 0.45 & 0.32 & -1.19 & 0.50 & 0.37 \\
-12.97 & -0.89 & 0.45 & 0.37 & -1.00 & 0.53 & 0.42 & -1.14 & 0.56 & 0.52 \\
-12.63 & -0.84 & 0.55 & 0.48 & -0.96 & 0.61 & 0.57 & -1.10 & 0.71 & 0.64 \\
-12.29 & -0.82 & 0.65 & 0.63 & -0.94 & 0.74 & 0.73 & -1.08 & 0.81 & 0.87 \\
-11.95 & -0.81 & 0.76 & 0.83 & -0.93 & 0.88 & 0.96 & -1.09 & 0.95 & 1.14 \\
-11.61 & -0.83 & 0.89 & 1.10 & -0.97 & 0.98 & 1.32 & -1.14 & 1.10 & 1.51 \\
-11.27 & -0.88 & 1.07 & 1.41 & -1.05 & 1.17 & 1.71 & -1.24 & 1.30 & 1.97 \\
-10.93 & -0.99 & 1.27 & 1.84 & -1.19 & 1.41 & 2.18 & -1.41 & 1.54 & 2.56 \\
-10.59 & -1.16 & 1.47 & 2.42 & -1.40 & 1.64 & 2.87 & -1.65 & 1.81 & 3.32 \\
-10.25 & -1.40 & 1.72 & 3.18 & -1.68 & 1.93 & 3.75 & -1.99 & 2.19 & 4.30 \\
-9.91 & -1.72 & 2.08 & 4.10 & -2.07 & 2.35 & 4.84 & -2.44 & 2.61 & 5.59 \\
            \hline
        \end{tabularx}
    \end{center}
\caption{ BDA determination of the UV LF at redshifts 6, 7 and 8. The values of LF are given in logarithmic scale: $\phi\ [\rm{log_{10}(M_{\odot}\ mag^{-1}\ Mpc^{-3})}]$, $\sigma_{sup}$ and $\sigma_{inf}$ the superior and inferior 68\% C.I. .}
\label{Tab:LF678}
\end{table*}

\renewcommand{\tabularxcolumn}[1]{>{\centering\arraybackslash}m{#1}}
\begin{table*}
    \begin{center}
        \begin{tabularx}{\textwidth}{XXXXXXXXXX}
            \hline
              & \multicolumn{3}{|c}{$z=9$} & \multicolumn{3}{|c}{$z=10$}  & \multicolumn{3}{|c|}{$z=12$} \\
              \hline
            $M_{\rm{UV}}$ & \multicolumn{1}{|c}{$\phi$} & $\sigma_{sup}$  & $\sigma_{inf}$ & \multicolumn{1}{|c}{$\phi$} & $\sigma_{sup}$  & $\sigma_{inf}$  & \multicolumn{1}{|c}{$\phi$} & $\sigma_{sup}$ & \multicolumn{1}{c|}{$\sigma_{inf}$}  \\
            \hline
-20.11 & -4.37 & 0.18 & 0.14 & -4.84 & 0.19 & 0.16 & -5.90 & 0.21 & 0.17 \\
-19.77 & -4.13 & 0.14 & 0.12 & -4.57 & 0.15 & 0.12 & -5.58 & 0.16 & 0.15 \\
-19.43 & -3.90 & 0.11 & 0.09 & -4.31 & 0.12 & 0.10 & -5.27 & 0.13 & 0.13 \\
-19.09 & -3.68 & 0.09 & 0.09 & -4.07 & 0.11 & 0.09 & -4.97 & 0.12 & 0.12 \\
-18.75 & -3.47 & 0.09 & 0.09 & -3.84 & 0.10 & 0.10 & -4.69 & 0.12 & 0.14 \\
-18.41 & -3.27 & 0.09 & 0.10 & -3.62 & 0.10 & 0.11 & -4.42 & 0.14 & 0.14 \\
-18.07 & -3.08 & 0.10 & 0.11 & -3.40 & 0.11 & 0.13 & -4.17 & 0.14 & 0.17 \\
-17.73 & -2.89 & 0.12 & 0.12 & -3.20 & 0.13 & 0.14 & -3.92 & 0.17 & 0.19 \\
-17.39 & -2.71 & 0.13 & 0.13 & -3.01 & 0.16 & 0.15 & -3.69 & 0.20 & 0.19 \\
-17.05 & -2.54 & 0.14 & 0.16 & -2.82 & 0.17 & 0.17 & -3.47 & 0.21 & 0.21 \\
-16.71 & -2.38 & 0.16 & 0.16 & -2.64 & 0.18 & 0.18 & -3.26 & 0.21 & 0.22 \\
-16.37 & -2.23 & 0.17 & 0.17 & -2.48 & 0.18 & 0.19 & -3.07 & 0.22 & 0.23 \\
-16.03 & -2.09 & 0.16 & 0.18 & -2.33 & 0.18 & 0.20 & -2.89 & 0.23 & 0.23 \\
-15.69 & -1.96 & 0.17 & 0.18 & -2.19 & 0.18 & 0.20 & -2.74 & 0.22 & 0.24 \\
-15.35 & -1.85 & 0.18 & 0.18 & -2.08 & 0.19 & 0.20 & -2.61 & 0.22 & 0.25 \\
-15.01 & -1.75 & 0.19 & 0.20 & -1.97 & 0.21 & 0.22 & -2.49 & 0.24 & 0.27 \\
-14.67 & -1.66 & 0.23 & 0.22 & -1.87 & 0.26 & 0.23 & -2.38 & 0.33 & 0.27 \\
-14.33 & -1.58 & 0.29 & 0.24 & -1.78 & 0.32 & 0.26 & -2.27 & 0.39 & 0.31 \\
-13.99 & -1.50 & 0.36 & 0.28 & -1.70 & 0.40 & 0.31 & -2.17 & 0.48 & 0.39 \\
-13.65 & -1.42 & 0.46 & 0.33 & -1.62 & 0.53 & 0.35 & -2.09 & 0.63 & 0.43 \\
-13.31 & -1.36 & 0.55 & 0.43 & -1.55 & 0.61 & 0.47 & -2.01 & 0.71 & 0.59 \\
-12.97 & -1.30 & 0.64 & 0.57 & -1.50 & 0.71 & 0.64 & -1.95 & 0.84 & 0.79 \\
-12.63 & -1.27 & 0.75 & 0.76 & -1.46 & 0.83 & 0.85 & -1.92 & 1.01 & 1.02 \\
-12.29 & -1.25 & 0.90 & 0.99 & -1.45 & 1.00 & 1.09 & -1.91 & 1.17 & 1.35 \\
-11.95 & -1.27 & 1.03 & 1.32 & -1.48 & 1.12 & 1.49 & -1.95 & 1.33 & 1.81 \\
-11.61 & -1.34 & 1.23 & 1.71 & -1.56 & 1.35 & 1.92 & -2.06 & 1.56 & 2.37 \\
-11.27 & -1.46 & 1.47 & 2.21 & -1.71 & 1.58 & 2.49 & -2.26 & 1.84 & 3.08 \\
-10.93 & -1.66 & 1.70 & 2.90 & -1.93 & 1.86 & 3.25 & -2.53 & 2.16 & 4.01 \\
-10.59 & -1.94 & 1.99 & 3.80 & -2.25 & 2.18 & 4.27 & -2.92 & 2.60 & 5.18 \\
-10.25 & -2.32 & 2.42 & 4.89 & -2.68 & 2.64 & 5.49 & -3.43 & 3.07 & 6.68 \\
-9.91 & -2.82 & 2.88 & 6.31 & -3.21 & 3.15 & 7.00 & -4.00 & 3.63 & 8.20 \\
            \hline
        \end{tabularx}
    \end{center}
\caption{ BDA determination of the UV LF at redshifts 9, 10 and 12. The values of LF are given in logarithmic scale: $\phi\ [\rm{log_{10}(M_{\odot}\ mag^{-1}\ Mpc^{-3})}]$, $\sigma_{sup}$ and $\sigma_{inf}$ the superior and inferior 68\% C.I. .}
\label{Tab:LF91012}
\end{table*}

\renewcommand{\tabularxcolumn}[1]{>{\centering\arraybackslash}m{#1}}
\begin{table}
    \begin{center}
        \begin{tabularx}{\columnwidth}{XXXX}
            \hline
            $M_{\rm{UV}}$ & $\phi$ & $\sigma_{sup}$ & $\sigma_{inf}$ \\
            \hline
-20.11 & -7.82 & 0.25 & 0.20 \\
-19.77 & -7.39 & 0.21 & 0.17 \\
-19.43 & -6.98 & 0.17 & 0.17 \\
-19.09 & -6.60 & 0.16 & 0.17 \\
-18.75 & -6.23 & 0.16 & 0.19 \\
-18.41 & -5.88 & 0.18 & 0.22 \\
-18.07 & -5.55 & 0.21 & 0.24 \\
-17.73 & -5.24 & 0.25 & 0.25 \\
-17.39 & -4.94 & 0.26 & 0.28 \\
-17.05 & -4.66 & 0.27 & 0.30 \\
-16.71 & -4.40 & 0.30 & 0.29 \\
-16.37 & -4.16 & 0.29 & 0.31 \\
-16.03 & -3.94 & 0.29 & 0.31 \\
-15.69 & -3.76 & 0.28 & 0.31 \\
-15.35 & -3.60 & 0.28 & 0.32 \\
-15.01 & -3.45 & 0.32 & 0.35 \\
-14.67 & -3.32 & 0.43 & 0.34 \\
-14.33 & -3.18 & 0.52 & 0.39 \\
-13.99 & -3.06 & 0.62 & 0.49 \\
-13.65 & -2.96 & 0.79 & 0.56 \\
-13.31 & -2.86 & 0.91 & 0.75 \\
-12.97 & -2.79 & 1.08 & 0.98 \\
-12.63 & -2.75 & 1.25 & 1.32 \\
-12.29 & -2.75 & 1.42 & 1.76 \\
-11.95 & -2.81 & 1.65 & 2.33 \\
-11.61 & -2.97 & 1.97 & 3.00 \\
-11.27 & -3.21 & 2.26 & 3.95 \\
-10.93 & -3.58 & 2.63 & 5.18 \\
-10.59 & -4.06 & 3.17 & 6.61 \\
-10.25 & -4.60 & 3.69 & 8.16 \\
-9.91 & -5.10 & 4.17 & 9.38 \\
            \hline
        \end{tabularx}
    \end{center}
\caption{ BDA determination of the UV LF at redshift 15. The values of LF are given in logarithmic scale: $\phi\ [\rm{log_{10}(M_{\odot}\ mag^{-1}\ Mpc^{-3})}]$, $\sigma_{sup}$ and $\sigma_{inf}$ the superior and inferior 68\% C.I. .}
\label{Tab:LF15}
\end{table}
%%%%%%%%%%%%%%%%%%%%%%%%%%%%%%%%%%%%%%%%%%%%%%%%%%
%%%%%%%%%%%%%%%%%%%%%%%%%%%%%%%%%%%%%%%%%%%%%%%%%%

% Don't change these lines
\bsp    % typesetting comment
\label{lastpage}
\end{document}